\documentclass[aps,prb,twocolumn,showpacs,groupaddress,floatfix]{revtex4}
\usepackage{graphicx,graphics}
\usepackage{dcolumn}
\usepackage{amsmath,amssymb,amsfonts}
\usepackage{latexsym,verbatim}
\usepackage{bm}
\usepackage{color}
\usepackage{ulem}
\usepackage[breaklinks=true,colorlinks,citecolor=blue,linkcolor=blue,urlcolor=blue]{hyperref}

\def\be{\begin{equation}}
\def\ee{\end{equation}}

\begin{document}
\title{The intrinsic charge and spin conductivities of doped graphene in the Fermi-Liquid regime}

\author{Alessandro Principi}
\email{principia@missouri.edu}
\affiliation{Department of Physics and Astronomy, University of Missouri, Columbia, Missouri 65211, USA}	
\author{Giovanni Vignale}
\affiliation{Department of Physics and Astronomy, University of Missouri, Columbia, Missouri 65211, USA}

\begin{abstract}
The experimental availability of ultra-high-mobility samples of graphene opens the possibility to realize and study experimentally the ``hydrodynamic'' regime of the electron liquid. In this regime the rate of electron-electron collisions is extremely high and dominates over the electron-impurity and electron-phonon scattering rates, which are therefore neglected. The system is brought to a local quasi-equilibrium described by a set of smoothly varying (in space and time) functions, {\it i.e.} the density, the velocity field and the local temperature.
In this paper we calculate the charge and spin conductivities of doped graphene due solely to electron-electron interactions. We show that, in spite of the linear low-energy band dispersion, graphene behaves in a wide range of temperatures as an effectively Galilean invariant system: the charge conductivity diverges in the limit $T \to 0$, while the spin conductivity remains finite. These results pave the way to the description of charge transport in graphene in terms of Navier-Stokes equations.
\end{abstract}
\maketitle

\section{Introduction}
Graphene, a two-dimensional (2D) layer of carbon atoms arranged in a honeycomb lattice, has attracted a huge deal of both theoretical and experimental interests in the last few years~\cite{castroneto_rmp_2009,Peres_RMP_2010,Bonaccorso_NatPhot_2010,Goerbig_RMP_2011,DasSarma_rmp_2011,kotov_rmp_2012}. Its properties, due to the gapless and linear low-energy band dispersion, make it highly attractive for several practical applications spanning the fields of optoelectronics, photonics, nanoplasmonics, metrology, and energy generation and storage~\cite{Bonaccorso_NatPhot_2010,Novoselov_nature_2012,Vakil_science_2011,Koppens_nanolett_2011,Tassin_natphot_2012,Bao_acsnano_2012,Grigorenko_natphot_2012}. At the same time it still offers an extremely interesting playground for fundamental science. As the quality of samples continues to improve~\cite{Mayorov_nanolett_2012}, new breakthroughs are expected. Many-body interactions are indeed expected to play a crucial role in the physics of ultra-high-mobility samples~\cite{kotov_rmp_2012}.

The transport properties of graphene are controlled by the highly-mobile electrons in the $\pi$ (valence) and $\pi^\star$ (conduction) bands, which arise from the hybridization of the $p_z$ orbitals~\cite{castroneto_rmp_2009} (${\hat {\bm z}}$ denotes the direction orthogonal to the graphene plane). The two bands touch with a linear dispersion at two inequivalent points ($K$ and $K'$) at the corners of the hexagonal Brillouin zone. For small energies, momenta, and doping concentrations, it is possible to expand the tight-binding Hamiltonian around these points in a ${\bm k}\cdot{\bm p}$ fashion~\cite{castroneto_rmp_2009}. The resulting low-energy Hamiltonian describes massless Dirac fermions (MDFs) characterized by the density-independent Fermi velocity $v_{\rm F}$ which is about $300$ times smaller than the speed of light.

A high-mobility gas of free carriers can be created in graphene by, e.g., electrostatic gating or chemical doping.  We assume these carriers to be in the ``Fermi-liquid'' regime~\cite{Giuliani_and_Vignale}.  The Fermi energy is thus $\varepsilon_{\rm F} = \pm v_{\rm F} k_{\rm F}$ (energies are measured from the Dirac point), where $k_{\rm F} = \sqrt{2\pi n/N_{\rm v}}$ is the Fermi wavevector, $n$ is the excess carrier density, and $N_{\rm v}=2$ is the valley degeneracy. The sign of $\varepsilon_{\rm F}$ depends on the type of carriers ($+$ for electrons and $-$ for holes). In what follows, owing to the particle-hole symmetry of the low-energy MDF model~\cite{castroneto_rmp_2009}, we consider only samples doped with an excess electron density.

Most of the previous theoretical works on the transport properties of doped graphene~\cite{DasSarma_rmp_2011} considered samples in which the transport is dominated by disorder effects, and in which electron-electron interactions play a minor or no role. In this paper instead, in view of the possibilities offered by the experimentally-available ultra-high-mobility samples~\cite{Mayorov_nanolett_2012}, we focus on the ``intrinsic'' transport regime. By intrinsic we mean that the transport properties are solely determined by electron-electron interactions, and disorder and phonons, as well as finite size effects, are considered to be irrelevant. In this Paper we focus on two fundamental properties of doped graphene, namely the charge and spin conductivities. Let us write these conductivities, at a finite frequency $\omega$,  in the common form
\begin{equation} \label{eq:conductivities_general}
\sigma^{(\ell)}(\omega)
=
\frac{Q_\ell^2 {\cal D}_\ell}{-i\omega + 1/\tau_{\rm tr}^{(\ell)}}
~,
\end{equation}
where  $\ell={\rm c}$ for the charge conductivity, $\ell= {\rm s}$ for the spin conductivity, $Q_{\rm c} = e$ and $Q_{\rm s} = \hbar$. Here $\tau_{\rm tr}^{({\rm c})}$ and $\tau_{\rm tr}^{({\rm s})}$ are the transport relaxation times of charge and spin currents, respectively, while ${\cal D}_{\rm c}$ and ${\cal D}_{\rm s}$ are the corresponding ``Drude weights''. 

Any conductivity, associated with the transport of a physical quantity, are, in general,  affected by electron-electron interactions. Both its ``Drude weight'' and its relaxation time change when interactions are turned on. 
In fact, as it happens for the quasiparticle lifetime ($\tau_{\rm ee}$), also transport relaxation times are limited by thermally-activated electron-electron scattering processes.
Moreover, interactions renormalize the quasiparticle properties as the Fermi velocity, and accordingly the Drude weights. The self-energy corrections which ``dress"  the interacting Fermi velocity, $v_{\rm F}^\star$, are especially important in graphene, since its low-energy MDF Hamiltonian has an infinite bandwidth~\cite{castroneto_rmp_2009,kotov_rmp_2012}. Interactions between electrons at the Fermi surface and states at large negative energies (in the valence band) lead to a logarithmic divergence of $v_{\rm F}^\star$~\cite{Gonzalez_nuclphys_1994} as the system approaches the undoped regime.
Furthermore, electron-electron interactions are also responsible for ``vertex corrections'' to the Drude weights, which are encoded in the Landau parameters~\cite{Giuliani_and_Vignale} $F_n^{{\rm a}/{\rm s}}$. Vertex corrections are usually small in a wide range of interaction strengths~\cite{Giuliani_and_Vignale}.
The full renormalization of the charge Drude weight takes the form~\cite{Giuliani_and_Vignale,abedinpour_prb_2011,Levitov_prb_2013} ${\cal D}_{\rm c} = {\cal D}_{\rm c}^{(0)} v_{\rm F}^\star(1+F_1^{\rm s})/v_{\rm F}$, while the spin Drude weight becomes~\cite{Qian_prl_2004} ${\cal D}_{\rm s} = {\cal D}_{\rm s}^{(0)} v_{\rm F}^\star(1+F_1^{\rm a})/v_{\rm F}$ (see also App.~\ref{eq:spin_velocity}).

Although this is a very general scheme, care must be exerted when dealing with charge currents. In the case of a parabolic band 2D electron gas (2DEG), for example, Galilean invariance leads to a perfect cancellation between self-energy and vertex corrections to the charge Drude weight. In formulas $v_{\rm F}^\star(1+F_1^{\rm s}) = v_{\rm F}$. For the same reason, the charge transport time of a clean Galilean-invariant Fermi liquid is unaffected by electron-electron interactions in the absence of Umklapp processes, and it is thus infinite as in the non-interacting case. These facts show that electron-electron interactions, which conserve the total momentum of the system at any scattering event, are also inefficient in relaxing a homogeneous current. In Galilean invariant systems, the latter is indeed proportional to the total momentum, which is a conserved quantity.  At odds with this, the spin current follows the ``general rule'' outlined above and is relaxed by electron-electron interactions, which can transfer momentum between the two spin populations, giving rise to the phenomenon of the spin-Coulomb drag~\cite{DAmico_prb_2000,DAmico_prb_2002,DAmico_prb_2003}.

Since graphene is not a Galilean invariant system, charge relaxation due to electron-electron interactions is not forbidden {\it a priori} by any symmetry. In the low-energy MDF model, indeed, the current and total momentum are not proportional to each other. It is therefore not surprising that electron-electron interactions affect the charge Drude weight of graphene in a non-trivial way~\cite{abedinpour_prb_2011,Levitov_prb_2013}, {\it i.e.} the product $v_{\rm F}^\star (1+F_1^{\rm s}) \neq v_{\rm F}$. It has been shown that, to the first order in the strength of electron-electron interactions~\cite{abedinpour_prb_2011}, vertex corrections exactly cancel the self-energy renormalization due to particle-particle scattering at the Fermi surface, but do not affect the logarithmic divergence of the Fermi velocity~\cite{abedinpour_prb_2011}. Moreover, while it is clear that in a 2DEG the charge transport time is unaffected by interactions, no conclusion can be drawn {\it a priori} for graphene.

In this paper we prove that at low temperature the charge transport time is infinite, {\it i.e.} $1/\tau_{\rm tr}^{(\rm c)} = 0^+$, while the spin transport time $\tau_{\rm tr}^{(\rm s)}$ is limited by electron-electron interaction and it is thus finite. As noted above, the relation between the current and total momentum is highly non-linear in graphene. Is it thus somewhat surprising that the charge transport time is not affected by electron-electron interactions, and that only the Drude weight is renormalized.
This result can be understood as follows. 

While the momentum ${\bm k}$ and velocity ${\bm v}_\lambda = \lambda v_{\rm F} {\bm k}/|{\bm k}|$ of a quasiparticle are not directly proportional to each other, they become approximately linearly related  at  low temperature for any finite doping concentration.  Indeed, in the limit of $k_{\rm B} T \ll \varepsilon_{\rm F}$, the dominant contribution to the transport comes from electrons in a thin shell of size $k_{\rm B} T$ around the Fermi energy. All these quasiparticles have magnitude of momentum equal to $k_{\rm F}$, and velocity ${\bm v}_+ \simeq v_{\rm F} {\bm k}/k_{\rm F}$, if the system is n-doped. This in turn implies that a linear relation is established between the momentum and velocity of each quasiparticle and, accordingly, between the current and total momentum of the system. Since the latter is conserved, at low temperature electron-electron interactions cannot relax a homogeneous current and doped graphene behaves as an effectively Galilean-invariant system.

We stress that this argument applies only to the calculation of the charge transport time and breaks down when one considers the charge Drude weight. The latter has contributions from virtual processes between all quasiparticle states, not only those around the Fermi energy. Since these processes span all the quasiparticle spectrum, the non-linear relation between the current and momentum operators becomes apparent and the Drude weight gets renormalized. 

The situation is completely different for the spin conductivity. In this case electron-electron interactions (i) renormalize the spin Drude weight and (ii) provide a finite transport time for spin currents. While the former effect is expected to be small in a wide range of values of the strength of electron-electron interactions~\cite{Qian_prl_2004}, the latter is large. The spin conductivity, which was infinite in the non-interacting limit, turns out to be finite in an interacting system. It thus offers a more powerful probe of electron-electron interactions as compared with the charge conductivity. The physics behind the spin conductivity is intimately related to the phenomenon of spin-Coulomb drag. When a pure spin-polarized current is injected into the system, each spin component of the current exerts friction on the other spin component via electron-electron interactions. The relative velocity of the spin populations therefore decays in time, and eventually vanishes unless an external driving field is present, in which case it reaches a steady state. This in turn implies that the spin conductivity must be finite. Our calculation shows that typical values for the spin transport time range between $1-10~{\rm ps}$. 

This paper is organized as follows. In Sect.~\ref{sect:model} we define the low-energy MDF model of graphene, and we set up the all-order diagrammatic calculations needed to determine the charge and spin conductivities. The main steps of the calculation are given in Sect.~\ref{sect:FL_conductivity}, which  also presents the main results of our paper, namely the charge and spin transport times.
Our results are summarized in Sect.~\ref{sect:summary}. Appendices.~\ref{app:quasiparticle_lifetime}-\ref{eq:spin_velocity} provide several technical details of the calculation.

\section{Model and basic definitions}
\label{sect:model}
We model graphene with the low-energy MDF Hamiltonian (per valley flavor -- hereafter $\hbar = 1$)~\cite{castroneto_rmp_2009,kotov_rmp_2012}
\begin{equation} \label{eq:MDF_Hamiltonian}
{\hat {\cal H}} = \sum_{{\bm k},\lambda} \varepsilon_{{\bm k},\lambda} {\hat \psi}^\dagger_{{\bm k},\lambda,\sigma} {\hat \psi}_{{\bm k},\lambda,\sigma}
+ \frac{1}{2} \sum_{{\bm q}} v_{\bm q} ({\hat n}_{{\bm q}} {\hat n}_{-{\bm q}} - {\hat n}_{{\bm 0}})
~,
\end{equation}
where $\psi_{{\bm k},\lambda,\sigma}$ ($\psi^\dagger_{{\bm k},\lambda,\sigma}$) destroys (creates) a particle with momentum ${\bm k}$ and spin $\sigma=\pm$ in band $\lambda=\pm$, $\varepsilon_{{\bm k},\lambda} = \lambda v_{\rm F} k$, and $v_{\bm q} = 2\pi e^2/(\epsilon q)$ is the non-relativistic Coulomb interaction. Here $\epsilon$ models the dielectric environment surrounding graphene and, as a first approximation, it is the average of the dielectric constants of media above ($\epsilon_1$) and below ($\epsilon_2$) the sheet, {\it i.e.} $\epsilon = (\epsilon_1 + \epsilon_2)/2$. The strength of electron-electron interactions is characterized by the density-independent ``fine-structure constant'' of graphene (restoring $\hbar$) $\alpha_{\rm ee} = e^2/(\hbar \epsilon v_{\rm F})$. Finally, the density operator is 
\begin{equation} \label{eq:density_op}
{\hat n}_{\bm q} = \!\! \sum_{{\bm k},\sigma,\lambda,\lambda'} {\cal D}_{\lambda\lambda'}({\bm k}_-, {\bm k}_+) {\hat \psi}^\dagger_{{\bm k}_-,\lambda,\sigma} {\hat \psi}_{{\bm k}_+,\lambda',\sigma}
~,
\end{equation}
where we defined ${\bm k}_\pm = {\bm k} \pm {\bm q}/2$, and the matrix element of the density operator between the eigenstates of the bare Hamiltonian is~\cite{castroneto_rmp_2009}
\begin{equation} \label{eq:D_element}
{\cal D}_{\lambda\lambda'} ({\bm k}, {\bm k}') =
\frac{e^{i(\varphi_{\bm k}-\varphi_{{\bm k}'})/2} + \lambda\lambda' e^{-i(\varphi_{\bm k}-\varphi_{{\bm k}'})/2}}{2}
~.
\end{equation}
Here $\varphi_{\bm k}$ is the angle between the momentum ${\bm k}$ and the ${\hat {\bm x}}$-axis. 

The spin-resolved current operator of MDFs is
\begin{equation} \label{eq:j_sigma_def}
{\hat {\bm j}}^{(\sigma)}_{{\bm q}} = \!\! \sum_{{\bm k},\lambda,\lambda'} {\bm J}_{\lambda\lambda'}({\bm k}_-, {\bm k}_+) {\hat \psi}^\dagger_{{\bm k}_-,\lambda,\sigma} {\hat \psi}_{{\bm k}_+,\lambda',\sigma}
~,
\end{equation}
where ${\bm J}_{\lambda\lambda'}({\bm k},{\bm k}') \equiv [J^{(x)}_{\lambda\lambda'}({\bm k},{\bm k}'), J^{(y)}_{\lambda\lambda'}({\bm k},{\bm k}')]$, and 
\begin{eqnarray} \label{eq:J_element}
J^{(x)}_{\lambda\lambda'} ({\bm k}, {\bm k}') &=&
v_{\rm F} \frac{\lambda' e^{i(\varphi_{\bm k}+\varphi_{{\bm k}'})/2} + \lambda e^{-i(\varphi_{\bm k}+\varphi_{{\bm k}'})/2}}{2}
~,
\nonumber\\
J^{(y)}_{\lambda\lambda'} ({\bm k}, {\bm k}') &=& 
v_{\rm F} \frac{\lambda' e^{i(\varphi_{\bm k}+\varphi_{{\bm k}'})/2} - \lambda e^{-i(\varphi_{\bm k}+\varphi_{{\bm k}'})/2}}{2i}
~,
\end{eqnarray}
are the matrix elements of the current operator between the eigenstates $|{\bm k},\lambda\rangle$ and $|{\bm k}',\lambda'\rangle$ of the system.

\begin{figure}[t]
\begin{center}
\begin{tabular}{c}
\includegraphics[width=0.99\columnwidth]{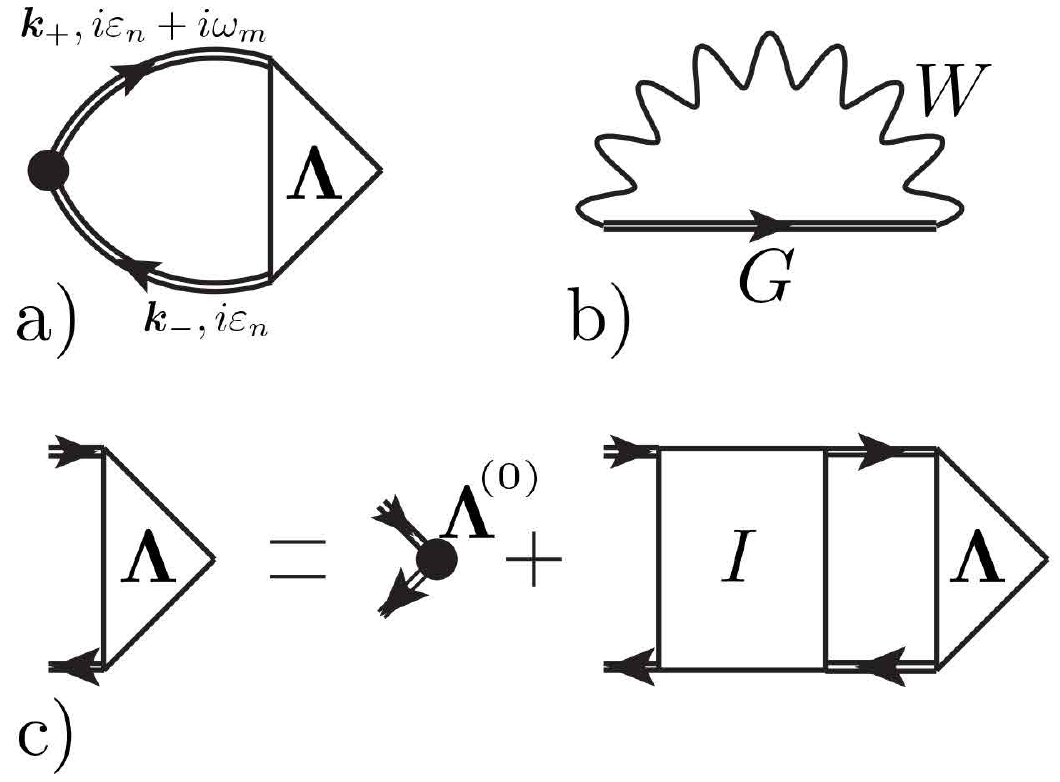}
\end{tabular}
\end{center}
\caption{
a) The diagrammatic representation of the current-current response function. The left dot is the bare vertex $\Lambda^{(0,\alpha)}$ (we suppress the momentum-energy dependence for brevity), while the solid double lines are Green's functions dressed by the self-energy. In the large-$n_{\rm F}$ limit it correspond to the $GW$ self-energy, which is depicted in panel b). Wavy lines represent RPA screened interactions. Finally, the triangle represents the vertex function $\Lambda^{\beta}$ which is dressed by e-e interactions and satisfies the Bethe-Salpeter equation in panel c). Note that the form of the irreducible interaction $I$ is uniquely determined by the choice of the self-energy, provided $\Lambda^\beta$ must satisfy the Ward identities~\cite{Giuliani_and_Vignale} (see Fig.~\ref{fig:two}).
\label{fig:one}}
\end{figure}

\begin{figure}[t]
\begin{center}
\begin{tabular}{c}
\includegraphics[width=0.99\columnwidth]{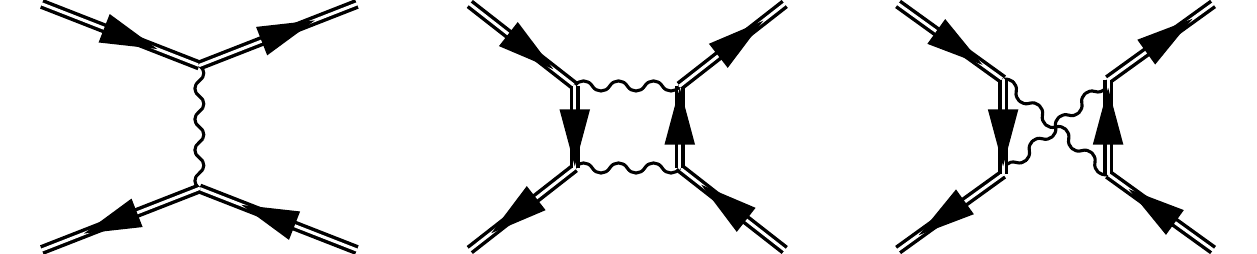}
\end{tabular}
\end{center}
\caption{
The diagrams that contribute to the irreducible interaction $I$ of Fig.~\ref{fig:one}.
\label{fig:two}}
\end{figure}

Fig.~\ref{fig:one} summarizes the all-order diagrammatic re-summation needed to calculate the charge and spin conductivities,~\cite{Eliashberg_jetp_1962,Yamada_ptp_1986} which are microscopically defined as~\cite{proper_chi}
\begin{equation} \label{eq:sigma_c_def}
\sigma^{({\rm c})}_{\alpha\beta} = \lim_{\omega\to 0}\left[\frac{i e^2}{\omega} \sum_{\sigma\sigma'} \chi_{j_{\alpha}^{(\sigma)} j_{\beta}^{(\sigma')}} ({\bm q}={\bm 0}, \omega)\right]
~,
\end{equation}
and
\begin{equation} \label{eq:sigma_s_def}
\sigma^{({\rm s})}_{\alpha\beta} = \lim_{\omega\to 0}\left[\frac{i \hbar^2}{\omega} \sum_{\sigma\sigma'} \sigma\sigma' \chi_{j_{\alpha}^{(\sigma)} j_{\beta}^{(\sigma')}} ({\bm q}={\bm 0}, \omega)\right]
~.
\end{equation}
Here $\chi_{j_{\alpha}^{(\sigma)} j_{\beta}^{(\sigma')}} ({\bm q}={\bm 0}, \omega)$ is the proper spin-resolved current-current linear response function~\cite{Giuliani_and_Vignale} given by the diagram in Fig.~\ref{fig:one}a). Its analytical expression on the imaginary-frequency axis reads
\begin{eqnarray} \label{eq:chi_jj_def}
&& \!\!\!\!\!\!\!\!
\chi_{j_{\alpha}^{(\sigma)} j_{\beta}^{(\sigma')}} ({\bm q}, i\omega_m) = N_{\rm v} k_{\rm B} T \sum_{{\bm k}, \lambda, \lambda'} \sum_{\varepsilon_n} G^{(\sigma)}_{\lambda}({\bm k}_-, i\varepsilon_n) 
\nonumber\\
&\times&
\Lambda^{(0,\sigma)}_{\lambda\lambda',\alpha}({\bm k}_-,{\bm k}_+) G^{(\sigma)}_{\lambda'}({\bm k}_+, i\varepsilon_n + i\omega_m)
\nonumber\\
&\times&
\Lambda^{(\sigma\sigma')}_{\lambda'\lambda,\beta} ({\bm k}_+, i\varepsilon_n + i\omega_m,{\bm k}_-,i\varepsilon_n)
~.
\end{eqnarray}
Here $\alpha, \beta = x, y$ denote the Cartesian components of the vectors. The double lines in Fig.~\ref{fig:one} correspond to the spin-resolved Green's functions which are dressed by the GW self-energy insertion of Fig.~\ref{fig:one}b) (see Sect.~\ref{sect:self_energy} below). In Eq.~(\ref{eq:chi_jj_def}) $\varepsilon_n = (2n+1)/\beta$ ($\omega_m = 2m/\beta$) are fermionic (bosonic) Matsubara frequencies ($n,m=0,1,\ldots$), $G^{(\sigma)}_{\lambda}({\bm k}, i\varepsilon_n)$ is the Green's function on the imaginary-frequency axis, while the bare current vertex determined from Eq.~(\ref{eq:j_sigma_def}) reads
\begin{equation} \label{eq:bare_vertex}
\Lambda^{(0,\sigma)}_{\lambda\lambda',\alpha}({\bm k}_-,{\bm k}_+) =
J^{(\alpha)}_{\lambda\lambda'} ({\bm k}_-, {\bm k}_+)
~.
\end{equation}
The term in the last line of Eq.~(\ref{eq:chi_jj_def}), namely $\Lambda^{(\sigma\sigma')}_{\lambda'\lambda,\beta} ({\bm k}_+, i\varepsilon_n + i\omega_m,{\bm k}_-,i\varepsilon_n)$, is the vertex function, which is dressed by electron-electron interactions and satisfies the self-consistent Bethe-Salpeter equation of Fig.~\ref{fig:one}c) (see Sect.~\ref{sect:bethe_salpeter} below). We stress that the choice of the self-energy, together with the requirement of fulfilling the Ward identities, uniquely determines the self-consistent Bethe-Salpeter equation satisfied by the vertex function, {\it i.e.} the irreducible interaction $I$.~\cite{Giuliani_and_Vignale}

The $GW$ self-energy shown in Fig.~\ref{fig:one}b) reads
\begin{eqnarray} \label{eq:GW}
\Sigma^{(\sigma)}_\lambda({\bm k},i\varepsilon_n) &=& -k_{\rm B} T \sum_{{\bm k}',\lambda'} \sum_{\varepsilon_{n'}} W({\bm k}'-{\bm k},i\varepsilon_{n'} - i\varepsilon_n)
\nonumber\\
&\times&
G_{\lambda'}^{(\sigma)}({\bm k}',i\varepsilon_{n'}) {\cal D}_{\lambda\lambda'}({\bm k},{\bm k}') {\cal D}_{\lambda'\lambda}({\bm k}',{\bm k})
~.
\nonumber\\
\end{eqnarray}
Here $W({\bm q},i\Omega_{m})$ is the screened electron-electron interaction, represented in Fig.~\ref{fig:one}b) by a wavy line. In the large-$N_{\rm v}$ limit this is given by
\begin{equation} \label{eq:W_RPA}
W({\bm q},i\Omega_{m}) = \frac{v_{\bm q}}{1-v_{\bm q} \chi_{nn}({\bm q},i\Omega_m)}
~,
\end{equation}
where $\chi_{nn}({\bm q},\omega)$ is the proper density-density response function~\cite{Giuliani_and_Vignale} of graphene. In principle, this should be calculated in analogy to the current-current response function of Eq.~(\ref{eq:chi_jj_def}), {\it i.e.} it should contain dressed Green's functions and vertex function. However, to simplify our calculation we neglect the vertex corrections to $\chi_{nn}({\bm q},\omega)$, which we define as
\begin{eqnarray} \label{eq:chi_nn}
\chi_{nn}({\bm q},i\omega_m) &=& N_{\rm v} k_{\rm B} T \sum_{{\bm q}',\varepsilon_n,\sigma'} \sum_{\lambda'',\mu''} G_{\lambda''}^{(\sigma')}({\bm q}',i\varepsilon_n)
\nonumber\\
&\times&
G_{\mu''}^{(\sigma')}({\bm q'}+{\bm q}, i\varepsilon_n+i\omega_m)
\nonumber\\
&\times&
{\cal D}_{\lambda''\mu''}({\bm q}',{\bm q}'+{\bm q}) {\cal D}_{\mu''\lambda''}({\bm q}'+{\bm q},{\bm q}') 
~.
\nonumber\\
\end{eqnarray}
Note that the density vertices in Eq.~(\ref{eq:chi_nn}) are {\it not} renormalized by electron-electron interactions.
This approximation does not invalidate the results obtained from Eq.~(\ref{eq:chi_jj_def}), {\it i.e.} it does not lead to qualitatively wrong behaviors of the conductivities. This because both self-energy and vertex corrections are taken into account on an equal footing in Eq.~(\ref{eq:chi_jj_def}). The approximation done in Eq.~(\ref{eq:chi_nn}) has however an impact on the form of the Bethe-Salpeter equation satisfied by $\Lambda^{(\sigma\sigma')}_{\lambda'\lambda,\beta} ({\bm k}_+, i\varepsilon_n + i\omega_m,{\bm k}_-,i\varepsilon_n)$, {\it i.e.} on the form of the irreducible interaction $I$ [see Fig.~\ref{fig:two}]. 
Indeed, the requirement of fulfilling the Ward identities~\cite{Giuliani_and_Vignale} constrains the dressed vertex to satisfy the following self-consistent Bethe-Salpeter equation~\cite{Giuliani_and_Vignale}:
\begin{eqnarray} \label{eq:Lambda_def}
&& \!\!\!\!\!\!\!\!
\Lambda^{(\sigma\sigma')}_{\lambda'\lambda,\beta} ({\bm k}_+, i\varepsilon_n + i\omega_m,{\bm k}_-,i\varepsilon_n) = \delta_{\sigma\sigma'}
\Lambda^{(0,\sigma)}_{\lambda'\lambda,\beta} ({\bm k}_+,{\bm k}_-)
\nonumber\\
&+&
\sum_{i=1,\ldots,3} \Lambda^{(i,\sigma\sigma')}_{\lambda'\lambda,\beta} ({\bm k}_+, i\varepsilon_n + i\omega_m,{\bm k}_-,i\varepsilon_n)
~.
\end{eqnarray}
The three contributions $\Lambda^{(i,\sigma\sigma')}_{\lambda'\lambda,\beta} ({\bm k}_+, i\varepsilon_n + i\omega_m,{\bm k}_-,i\varepsilon_n)$ (with $i=1,\ldots,3$) correspond to the three diagrams in Fig.~\ref{fig:two}. They read
\begin{eqnarray} \label{eq:Lambda_1_2}
&&
\Lambda^{(1,2,\sigma\sigma')}_{\lambda'\lambda,\beta} ({\bm k}_+, i\varepsilon_n + i\omega_m,{\bm k}_-,i\varepsilon_n) =
-k_{\rm B} T \sum_{{\bm k}',\varepsilon_{n'}} \sum_{\mu,\mu',\sigma''} 
\nonumber\\
&& \times
W^{(1,2,\sigma\sigma'')}_{\lambda\lambda'\mu\mu'}({\bm k}',{\bm k},i\varepsilon_{n'}-i\varepsilon_n)
G_{\mu'}^{(\sigma'')}({\bm k}'_+,i\varepsilon_{n'}+i\omega_m)
\nonumber\\
&& \times
G_{\mu}^{(\sigma'')}({\bm k}'_-,i\varepsilon_{n'})
\Lambda^{(\sigma''\sigma')}_{\mu'\mu,\beta} ({\bm k}'_+, i\varepsilon_{n'} + i\omega_m,{\bm k}'_-,i\varepsilon_{n'})
~,
\nonumber\\
\end{eqnarray}
and
\begin{eqnarray} \label{eq:Lambda_3}
&&
\Lambda^{(3,\sigma\sigma')}_{\lambda'\lambda,\beta} ({\bm k}_+, i\varepsilon_n + i\omega_m,{\bm k}_-,i\varepsilon_n) = 
-k_{\rm B} T \sum_{{\bm k}',\varepsilon_{n'}} 
\nonumber\\
&& \times
\sum_{\mu,\mu',\sigma''}
W^{(3,\sigma\sigma'')}_{\lambda\lambda'\mu\mu'}({\bm k}', {\bm k}, i\varepsilon_{n'}+i\varepsilon_{n} +i\omega_m)
\nonumber\\
&& \times
G_{\mu'}^{(\sigma'')}({\bm k}'_+,i\varepsilon_{n'}+i\omega_m) 
G_{\mu}^{(\sigma'')}({\bm k}'_-,i\varepsilon_{n'})
\nonumber\\
&& \times
\Lambda^{(\sigma''\sigma')}_{\mu'\mu,\beta} ({\bm k}'_+, i\varepsilon_{n'} + i\omega_m,{\bm k}'_-,i\varepsilon_{n'})
~.
\end{eqnarray}
Here we define
\begin{eqnarray} \label{eq:W_1}
&&
W^{(1,\sigma\sigma'')}_{\lambda\lambda'\mu\mu'} ({\bm k}',{\bm k},i\omega_m) = \delta_{\sigma\sigma''} W({\bm k}-{\bm k}',i\omega_m) 
\nonumber\\
&&\times
{\cal D}_{\lambda'\mu'}({\bm k}_+,{\bm k}'_+) {\cal D}_{\mu\lambda}({\bm k}'_-,{\bm k}_-)
~,
\end{eqnarray}
and
\begin{eqnarray} \label{eq:W_2}
&&
W^{(2,\sigma\sigma'')}_{\lambda\lambda'\mu\mu'} ({\bm k}',{\bm k},i\varepsilon_{n'}-i\varepsilon_n) =
N_{\rm v} k_{\rm B} T \sum_{{\bm q}',\omega_{m'}} 
\nonumber\\
&&\times
\sum_{\lambda'',\mu''} 
W({\bm q}',i\omega_{m'}) W({\bm q}'-{\bm q},i\omega_{m'}-i\omega_m) 
\nonumber\\
&& \times
{\cal D}_{\lambda'\lambda''}({\bm k}_+,{\bm k}_+-{\bm q}')
{\cal D}_{\lambda''\lambda}({\bm k}_+-{\bm q}',{\bm k}_-)
\nonumber\\
&& \times
{\cal D}_{\mu\mu''}({\bm k}'_-,{\bm k}'_+-{\bm q}')
{\cal D}_{\mu''\mu'}({\bm k}'_+-{\bm q}',{\bm k}'_+)
\nonumber\\
&& \times
G_{\lambda''}^{(\sigma)}({\bm k}_+-{\bm q}',i\varepsilon_n+i\omega_m -i\omega_{m'})
\nonumber\\
&& \times
G_{\mu''}^{(\sigma'')}({\bm k}'_+-{\bm q}',i\varepsilon_{n'}+i\omega_m -i\omega_{m'})
~,
\end{eqnarray}
and finally
\begin{eqnarray} \label{eq:W_3}
&&
W^{(3,\sigma\sigma'')}_{\lambda\lambda'\mu\mu'} ({\bm k}',{\bm k},i\varepsilon_{n'}+i\varepsilon_n+i\omega_m) =
N_{\rm v} k_{\rm B} T \sum_{{\bm q}',\omega_{m'}}
\nonumber\\
&& \times
\sum_{\lambda'',\mu''} 
W({\bm q}',i\omega_{m'}) W({\bm q}'-{\bm q},i\omega_{m'}-i\omega_m) 
\nonumber\\
&& \times
{\cal D}_{\lambda\lambda''}({\bm k}_-,{\bm k}_-+{\bm q}')
{\cal D}_{\lambda''\lambda'}({\bm k}_-+{\bm q}',{\bm k}_+)
\nonumber\\
&& \times
{\cal D}_{\mu\mu''}({\bm k}'_-,{\bm k}'_+-{\bm q}')
{\cal D}_{\mu''\mu'}({\bm k}'_+-{\bm q}',{\bm k}'_+)
\nonumber\\
&& \times
G_{\lambda''}^{(\sigma)}({\bm k}_-+{\bm q}',i\varepsilon_n +i\omega_{m'})
\nonumber\\
&& \times
G_{\mu''}^{(\sigma'')}({\bm k}'_+-{\bm q}',i\varepsilon_{n'}+i\omega_m -i\omega_{m'})
~.
\end{eqnarray}
In what follows we start from the evaluation of the self-energy corrections, and of the quasiparticle lifetime at the Fermi surface, and we then proceed to the calculation of the vertex correction.

\subsection{The quasiparticle decay rate} 
\label{sect:self_energy}
\begin{figure}[t]
\begin{center}
\begin{tabular}{c}
\includegraphics[width=0.6\columnwidth]{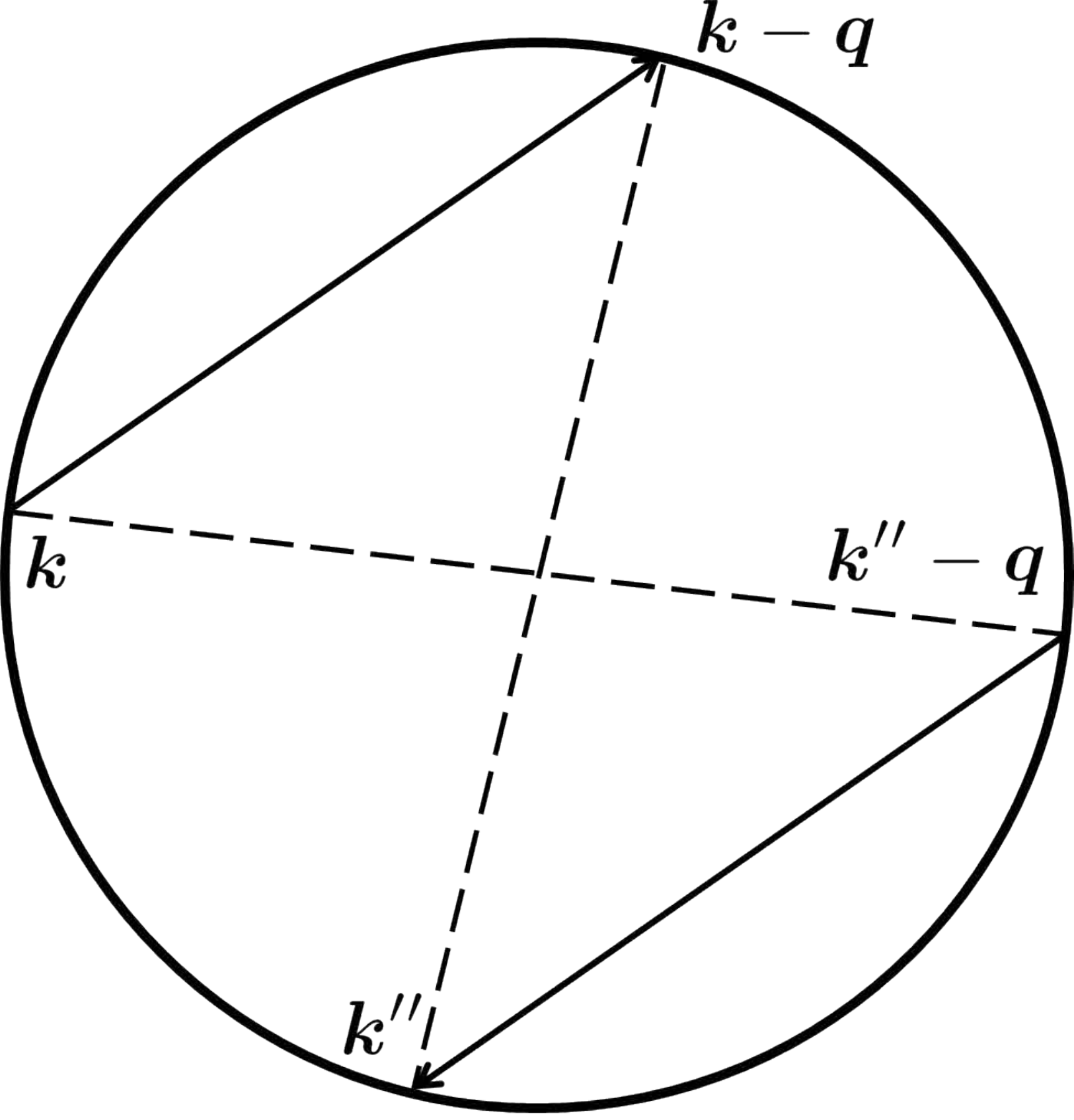}
\end{tabular}
\end{center}
\caption{
A pictorial representation of double particle-hole excitations that contribute, to lowest order in the strength of e-e interactions, to the quasiparticle decay rate calculated in Sect.~\ref{sect:self_energy}. Note that, since all the states involved in the scattering process live at the Fermi surface, the conservation of momentum constrains the initial states ${\bm k}$ and ${\bm k}'' - {\bm q}$ to be diametrically opposed. The same happens to the final states ${\bm k}-{\bm q}$ and ${\bm k}''$.
\label{fig:three}}
\end{figure}

In this section we calculate the quasiparticle lifetime $\tau_{\rm ee}$ at the Fermi energy, defined as
\begin{eqnarray} \label{eq:qp_decay_rate_def}
\frac{1}{\tau_{\rm ee}} &=& 2 \int_{-\infty}^{\infty} d\varepsilon \frac{\partial n_{\rm F}(\varepsilon)}{\partial \varepsilon} \Im m \big[\Sigma_+^{(\sigma)}({\bm k},\varepsilon^+)\big]\Big|_{k=k_{\rm F}}
~,
\end{eqnarray}
where the self-energy was defined in Eq.~(\ref{eq:GW}).
We focus on the imaginary part of the self-energy, which controls the charge and spin relaxation times, and we disregard its real part, which is responsible for the renormalization of the Drude weights. In the spirit of Landau theory of normal Fermi liquid, we take care of this approximation of the diagrammatic calculation by replacing {\it a posteriori} the non-interacting Drude weights with their interacting values. A microscopic calculation of the charge Drude weight to the first order in the strength of electron-electron interactions was given in Refs.~\onlinecite{abedinpour_prb_2011}.
As shown in App.~\ref{app:quasiparticle_lifetime}, at low temperature ($k_{\rm B}T\ll \varepsilon_{\rm F}$) Eq.~(\ref{eq:qp_decay_rate_def}) becomes
\begin{eqnarray} \label{eq:inverse_lifetime_3}
\frac{1}{\tau_{\rm qp}^{\rm ee}} 
&=&
\frac{(k_{\rm B} T)^2}{3 \pi} \int_{k_{\rm F}{\bar T}}^{2 k_{\rm F}(1-{\bar T}/2)} dq~q
|W({\bm q},{\bar T}\varepsilon_{\rm F})|^2
\nonumber\\
&\times&
\frac{\Im m \chi_{nn}({\bm q}, {\bar T}\varepsilon_{\rm F})}{{\bar T}\varepsilon_{\rm F}}
A(q,-{\bar T}\varepsilon_{\rm F})
~,
\nonumber\\
\end{eqnarray}
where ${\bar T} = \zeta k_{\rm B} T/\varepsilon_{\rm F}$, $\zeta= \pi/\sqrt{5}$, and 
\begin{eqnarray}
A(q,\omega) &=& -\frac{\pi}{v_{\rm F}} \int_0^{2\pi} d\varphi_{\bm q} \delta(|{\bm k}-{\bm q}| - k - \omega/v_{\rm F}) 
\nonumber\\
&\times&
\frac{1+\cos(\varphi_{{\bm k}-{\bm q}}+\varphi_{{\bm k}})}{2} \Bigg|_{k=k_{\rm F}}
\nonumber\\
&=&
-\frac{4\pi}{v_{\rm F}^2} \frac{k_{\rm F}-\omega/v_{\rm F}}{\sqrt{4 k_{\rm F}^2 q^2 - (q^2 - \omega^2/v_{\rm F}^2 + 2 k_{\rm F} \omega/v_{\rm F})^2}}
\nonumber\\
&\times&
\left( 1 - \frac{q^2-\omega^2/v_{\rm F}^2}{4 k_{\rm F}(k_{\rm F} - \omega/v_{\rm F})} \right)
\nonumber\\
&\times&
\Theta\left(1 - \left| \frac{q^2 - \omega^2/v_{\rm F}^2 + 2 k_{\rm F} \omega/v_{\rm F}}{2k_{\rm F}q} \right|\right)
~.
\end{eqnarray}
Eq.~(\ref{eq:inverse_lifetime_3}) describes, as shown in Fig.~\ref{fig:three}, the decay (scattering) of a quasiparticle of momentum ${\bm k}$ to a state of momentum ${\bm k}-{\bm q}$  through the creation of an electron-hole pair of total momentum ${\bm q}$ obtained by exciting a particle of momentum ${\bm k}'' - {\bm q}$ to a state of momentum ${\bm k}'' $.  Such a process is encoded in the density-density response function~\cite{DiracplasmonsRPA_1,DiracplasmonsRPA_2,DiracplasmonsRPA_3,DiracplasmonsRPA_4,DiracplasmonsRPA_5} $\Im m \chi_{nn}({\bm q}, {\bar T}\varepsilon_{\rm F})$ and is depicted in Fig.~\ref{fig:three}. Notice that, since all the initial and final states are on the Fermi surface, the conservation of momentum implies that ${\bm k}$ and ${\bm k}'' - {\bm q}$ (and thus ${\bm k}-{\bm q}$ and ${\bm k}''$) are diametrically opposite. This fact will be used in what follows to simplify the expressions of the transport times (see App~\ref{app:spin_transport_time}).

\begin{figure}[t]
\begin{center}
\begin{tabular}{c}
\includegraphics[width=0.99\columnwidth]{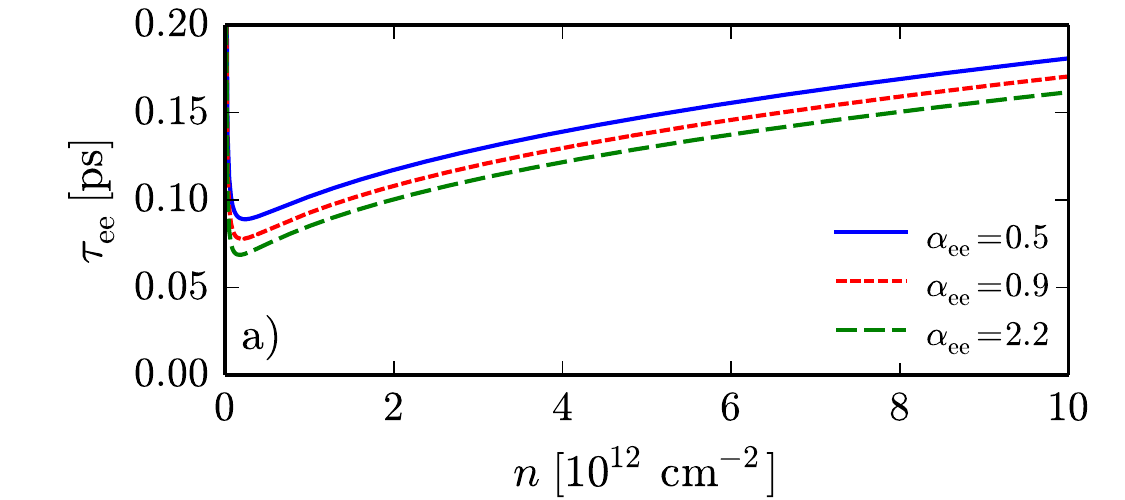}
\\
\includegraphics[width=0.99\columnwidth]{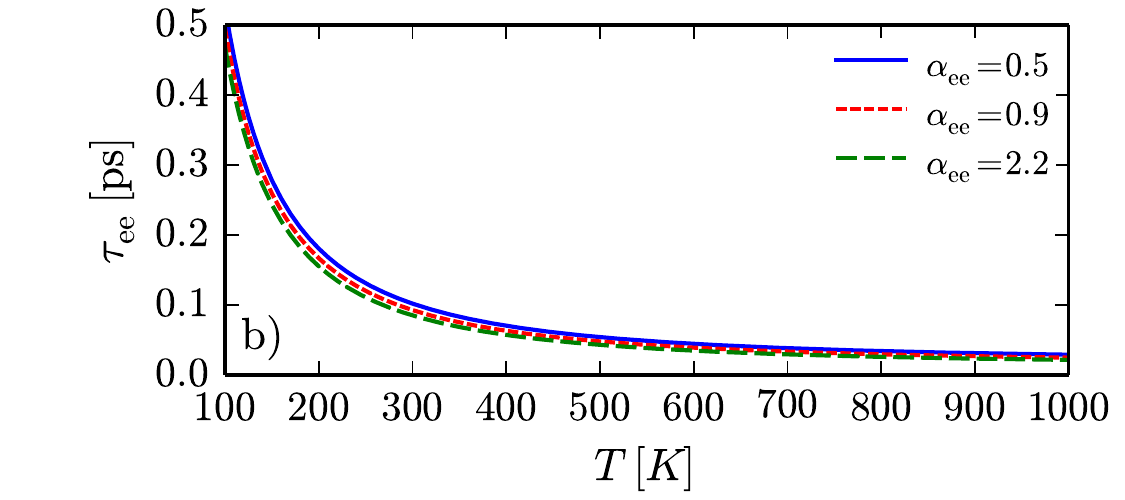}
\end{tabular}
\end{center}
\caption{
(Color online) Panel a) the quasiparticle lifetime of massless Dirac fermions $\tau_{\rm ee}$, as defined in Eq.~(\ref{eq:inverse_lifetime_3}), in units of picoseconds and plotted as a function of the density $n$ in units of $10^{12}~{\rm cm}^{-2}$ for three values of the dimensionless coupling constant $\alpha_{\rm ee}$. In this plot we fixed the temperature $T = 300~{\rm K}$. Panel b) same as in panel a) but shown as a function of temperature (in units of ${\rm K}$) for a fixed excess carrier density $n=10^{12}~{\rm cm}^{-2}$.
\label{fig:four}}
\end{figure}

\begin{figure}[t]
\begin{center}
\begin{tabular}{c}
\includegraphics[width=0.99\columnwidth]{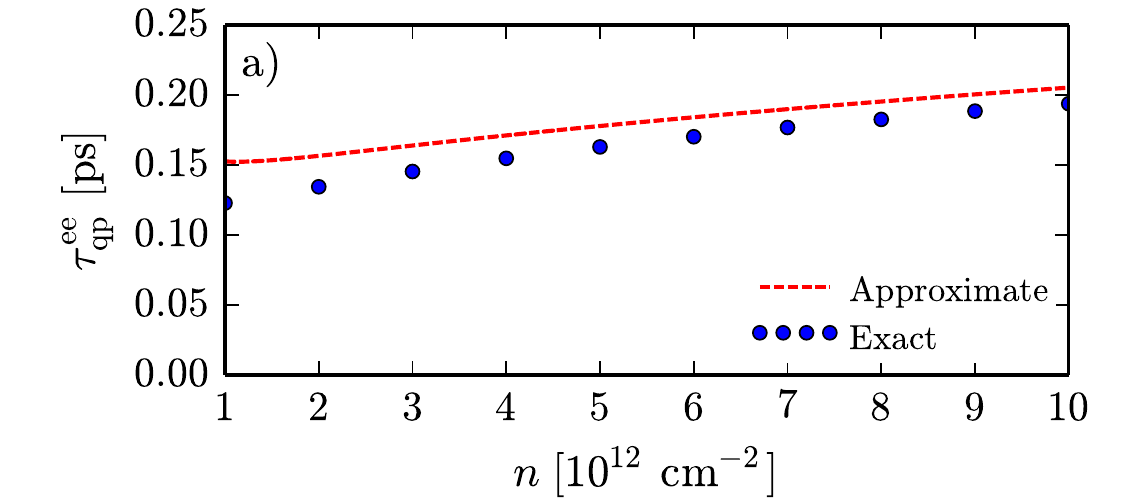}
\\
\includegraphics[width=0.99\columnwidth]{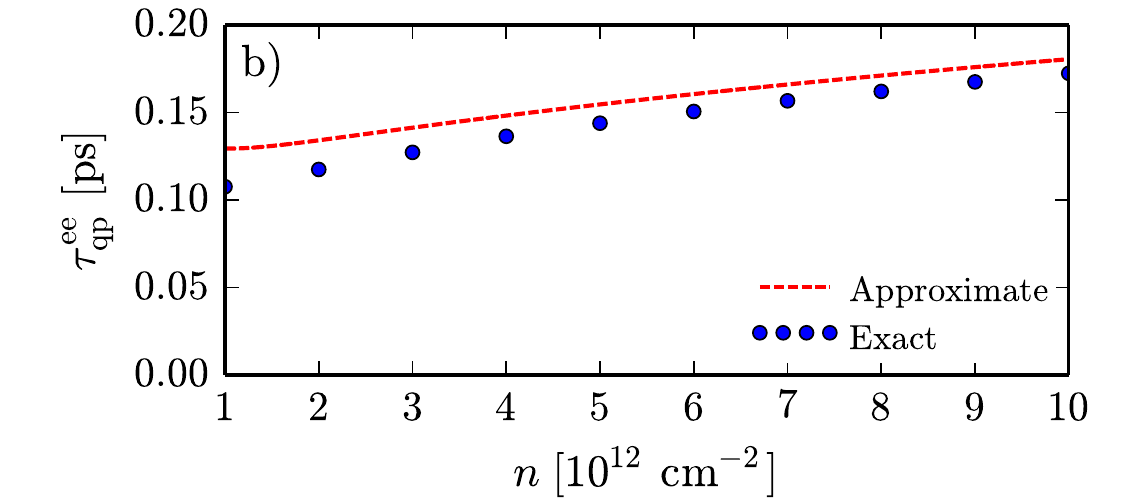}
\\
\includegraphics[width=0.99\columnwidth]{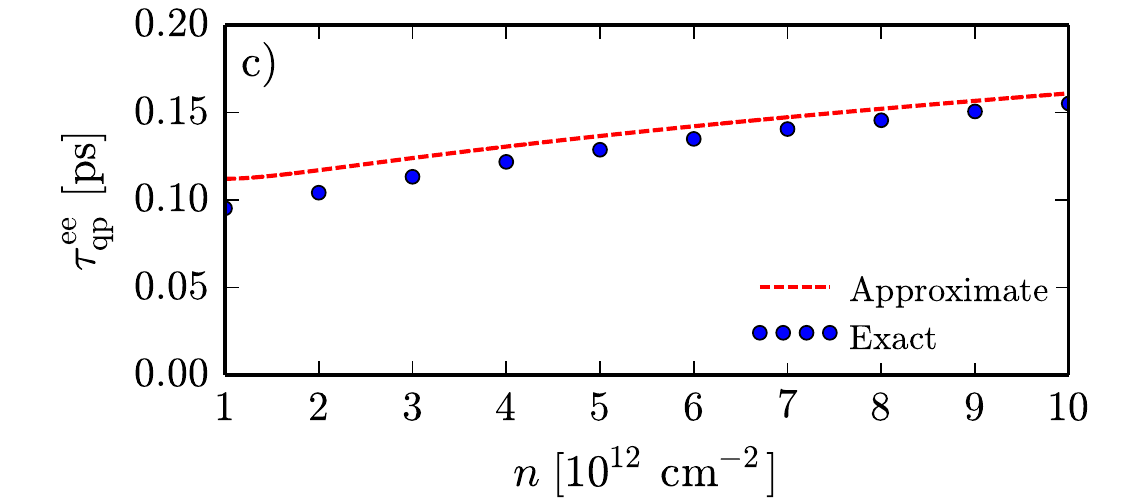}
\end{tabular}
\end{center}
\caption{
A comparison between the quasiparticle lifetime calculated from Eq.~(\ref{eq:inverse_lifetime_3}) and the one computed in Ref.~\onlinecite{Polini_QP_lifetime}. In this figure the temperature is kept fixed at $T=300~{\rm K}$ and the quasiparticle lifetime is plotted in units of ps as a function of the density (in units of $10^{12}~{\rm cm}^{-2}$). Panel a)-c) refer to the three values of the coupling constant $\alpha_{\rm ee}=0.5$, $\alpha_{\rm ee}=0.9$, and $\alpha_{\rm ee}=2.2$, respectively.
\label{fig:five}}
\end{figure}

\begin{figure}[t]
\begin{center}
\begin{tabular}{c}
\includegraphics[width=0.99\columnwidth]{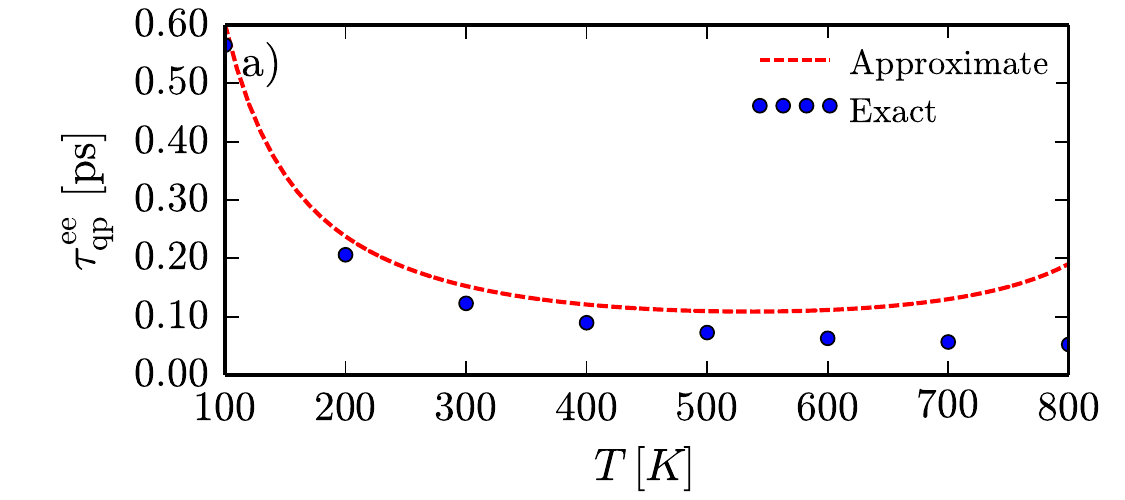}
\\
\includegraphics[width=0.99\columnwidth]{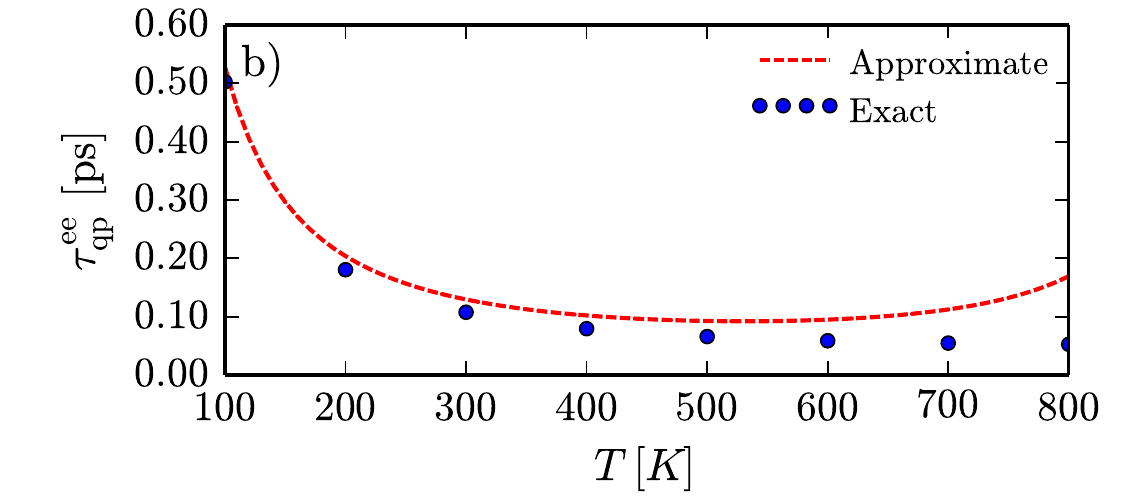}
\\
\includegraphics[width=0.99\columnwidth]{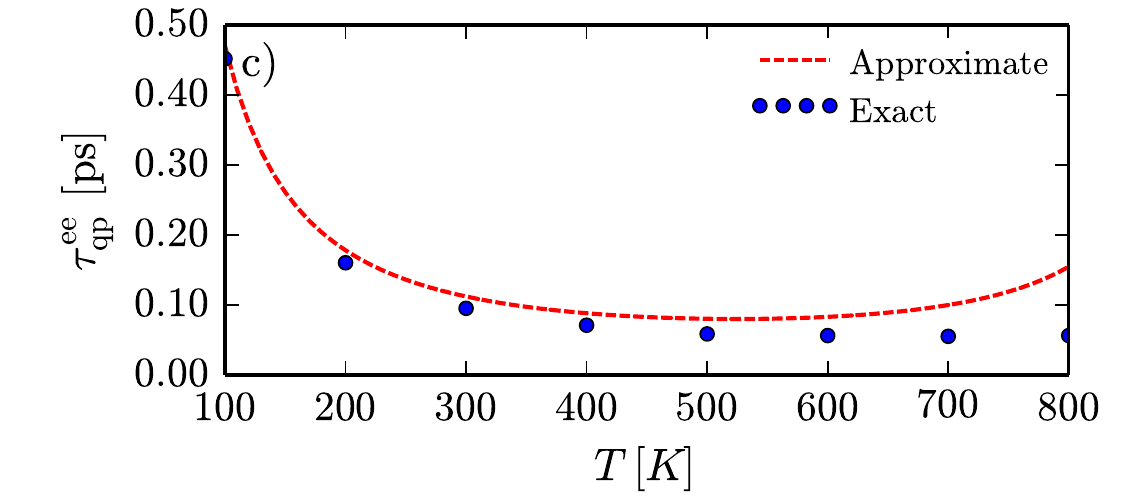}
\end{tabular}
\end{center}
\caption{
A comparison between the quasiparticle lifetime calculated from Eq.~(\ref{eq:inverse_lifetime_3}) and the one computed in Ref.~\onlinecite{Polini_QP_lifetime}. In this figure the density is kept fixed at $n = 10^{12}~{\rm cm}^{-2}$ and the quasiparticle lifetime is plotted in units of ps as a function of the temperature (measured in K). Panel a)-c) refer to the three values of the coupling constant $\alpha_{\rm ee}=0.5$, $\alpha_{\rm ee}=0.9$, and $\alpha_{\rm ee}=2.2$, respectively.
\label{fig:six}}
\end{figure}

Numerical results obtained from Eq.~(\ref{eq:inverse_lifetime_3}) are shown in Fig.~\ref{fig:four}. In passing, we recall that the GW quasiparticle lifetime is also calculated in Refs.~\onlinecite{Polini_QP_lifetime,Li_prb_2013}. Contrary to Eq.~(\ref{eq:inverse_lifetime_3}), the expressions given in Ref.~\onlinecite{Polini_QP_lifetime} do not contain any low-temperature approximation (see App.~\ref{app:quasiparticle_lifetime} for more details). 
In Figs.~\ref{fig:five}-\ref{fig:six} we show a comparison between the quasiparticle lifetime calculated from Eq.~(\ref{eq:inverse_lifetime_3}) and the ``exact'' one computed in Ref.~\onlinecite{Polini_QP_lifetime}. Note that the agreement is very good in the chosen range of temperatures and densities.

\subsection{The vertex function and the Bethe-Salpeter equation} 
\label{sect:bethe_salpeter}
In this section we summarize the complicated calculation of the vertex correction to the charge and spin conductivities. The details of the derivation can be found in Apps.~\ref{app:analytical_continuation} and~\ref{sect:fpm}.

The first step of the calculation is to analytically continue Eq.~(\ref{eq:chi_jj_def}) to real frequencies. We indeed recall that the charge and spin conductivities are determined, according to Eqs.~(\ref{eq:sigma_c_def}) and~(\ref{eq:sigma_s_def}), by the $\omega \to 0$ limit of the {\it retarded} spin-resolved current-current response function calculated at ${\bm q}={\bm 0}$. We stress that the analytical continuation $i\omega_m \to \omega+i\eta$ (with $\eta=0^+$) must be performed {\it before} the small-frequency limit.

After the analytical continuation to real frequencies~\cite{Bruus_and_Flensberg}, the current-current response function of Eq.~(\ref{eq:chi_jj_def}) contains products of advanced-advanced (schematically $G^{\rm A} G^{\rm A}$), retarded-retarded ($G^{\rm R} G^{\rm R}$) and advanced-retarded ($G^{\rm A} G^{\rm R}$) Green's function. The first two contributions, $G^{\rm A} G^{\rm A}$ and $G^{\rm R} G^{\rm R}$, have poles on the same side of the complex plane. In the limit $\varepsilon_{\rm F} \tau_{\rm ee} \gg 1$ we can neglect them~\cite{Bruus_and_Flensberg} and retain only the ``mixed'' term $G^{\rm A} G^{\rm R}$. After some simple algebra, shown in detail in App.~\ref{app:analytical_continuation}, Eq.~(\ref{eq:chi_jj_def}) becomes
\begin{eqnarray} \label{eq:R_chi_jj_final_omega_finite}
&& \!\!\!\!\!\!\!\!
\chi_{j_{\alpha}^{(\sigma)} j_{\beta}^{(\sigma')}}  ({\bm q}, \omega) = -N_{\rm v} \sum_{{\bm k}, \lambda, \lambda'} \int \frac{d\varepsilon}{2\pi i} \big[n_{\rm F} (\varepsilon+\omega) - n_{\rm F} (\varepsilon)\big]
\nonumber\\
&\times&
G^{({\rm A},\sigma)}_{\lambda} ({\bm k}_-, \varepsilon) \Lambda^{(0,\sigma)}_{\lambda\lambda',\alpha}({\bm k}_-,{\bm k}_+)
G^{({\rm R},\sigma')}_{\lambda'} ({\bm k}_+, \varepsilon + \omega)
\nonumber\\
&\times&
\Lambda^{(\sigma\sigma')}_{\lambda'\lambda,\beta} ({\bm k}_+, \varepsilon_+ + \omega,{\bm k}_-,\varepsilon_-)
~.
\end{eqnarray}

Accordingly, after the continuation to real frequencies the self-consistent Bethe-Salpeter Eq.~(\ref{eq:Lambda_def}) becomes
\begin{eqnarray} \label{eq:R_Lambda_def}
&& \!\!\!\!\!\!\!\!
\Lambda_{\lambda'\lambda,\beta}^{(\sigma\sigma')} ({\bm k}_+, \varepsilon_+ + \omega,{\bm k}_-,\varepsilon_-) = \delta_{\sigma\sigma'}
\Lambda^{(0,\sigma)}_{\lambda'\lambda,\beta} ({\bm k}_+,{\bm k}_-)
\nonumber\\
&+&
\sum_{i=1,\ldots,3} \Lambda^{(i,\sigma\sigma')}_{\lambda'\lambda,\beta} ({\bm k}_+, \varepsilon_+ + \omega,{\bm k}_-,\varepsilon_-)
~.
\end{eqnarray}
Also in the derivation of Eq.~(\ref{eq:R_Lambda_def}) we retained only mixed terms of the form $G^{\rm A} G^{\rm R}$. In spite of this approximation, our calculation yields a closed set of self-consistent equations. The terms $\{ \Lambda^{(i,\sigma\sigma')}_{\lambda'\lambda,\beta} ({\bm k}_+, \varepsilon_+ + \omega,{\bm k}_-,\varepsilon_-), i=1,\ldots,3 \}$ on the last line of Eq.~(\ref{eq:R_Lambda_def}) read
\begin{eqnarray} \label{eq:R_Lambda_1_together}
&&
\Lambda^{(1,\sigma\sigma')}_{\lambda'\lambda,\beta} ({\bm k}_+, \varepsilon_++\omega,{\bm k}_-,\varepsilon_-) =
4 N_{\rm v} \sum_{{\bm k}',{\bm q}',\sigma''} \sum_{\mu,\mu'} \sum_{\lambda'',\mu''} 
\nonumber\\
&& \times
\int \frac{d\varepsilon'}{2\pi i} \int \frac{d\omega'}{2\pi i} 
|W({\bm k}-{\bm k}',\varepsilon'-\varepsilon)|^2
\nonumber\\
&& \times
\big[n_{\rm F}(\varepsilon') + n_{\rm B}(\varepsilon'-\varepsilon)\big]
\big[n_{\rm F} (\omega'+\varepsilon') - n_{\rm F} (\omega'+\varepsilon)\big] 
\nonumber\\
&& \times
\Im m \Big[G^{({\rm R},\sigma'')}_{\lambda''}({\bm q}'-{\bm k},\omega'+\varepsilon)\Big]
\nonumber\\
&& \times
\Im m\Big[G^{({\rm R},\sigma'')}_{\mu''}({\bm q'}-{\bm k}', \omega'+\varepsilon')\Big]
\nonumber\\
&& \times
{\cal D}_{\lambda'\mu'}({\bm k}_+,{\bm k}'_+) {\cal D}_{\mu\lambda}({\bm k}'_-,{\bm k}_-)
\nonumber\\
&& \times
{\cal D}_{\lambda''\mu''}({\bm q}'-{\bm k},{\bm q}'-{\bm k}') {\cal D}_{\mu''\lambda''}({\bm q}'-{\bm k}',{\bm q}'-{\bm k}) 
\nonumber\\
&& \times
G^{({\rm R},\sigma)}_{\mu'}({\bm k}'_+,\varepsilon'+\omega) G^{({\rm A},\sigma)}_{\mu}({\bm k}'_-,\varepsilon')
\nonumber\\
&& \times
\Lambda^{(\sigma\sigma')}_{\mu'\mu,\beta} ({\bm k}'_+, \varepsilon'_++\omega, {\bm k}'_-,\varepsilon'_-)
~,
\end{eqnarray}
and
\begin{eqnarray} \label{eq:R_Lambda_2_together}
&&
\Lambda^{(2,\sigma\sigma')}_{\lambda'\lambda,\beta} ({\bm k}_+, \varepsilon_++\omega,{\bm k}_-,\varepsilon_-) =
4 N_{\rm v} \sum_{{\bm k}',{\bm q}',\sigma''} \sum_{\mu,\mu'} \sum_{\lambda'',\mu''} 
\nonumber\\
&& \times
\int \frac{d\varepsilon'}{2\pi i} \int \frac{d\omega'}{2\pi i}
W({\bm q}',\omega'_+) W({\bm q}',\omega'_--\omega) 
\nonumber\\
&& \times
\big[n_{\rm F}(\varepsilon') + n_{\rm B}(\varepsilon'-\varepsilon)\big]
\nonumber\\
&& \times
\big[n_{\rm F}(\omega'-\varepsilon-\omega) - n_{\rm F}(\omega'-\varepsilon'-\omega)\big]
\nonumber\\
&& \times
\Im m \big[ G^{({\rm R},\sigma)}_{\lambda''}({\bm k}_+-{\bm q}',\varepsilon+\omega-\omega') \big]
\nonumber\\
&& \times
\Im m \big[ G^{({\rm R},\sigma'')}_{\mu''}({\bm k}'_+-{\bm q}',\varepsilon'+\omega -\omega') \big]
\nonumber\\
&& \times
{\cal D}_{\lambda'\lambda''}({\bm k}_+,{\bm k}_+-{\bm q}')
{\cal D}_{\lambda''\lambda}({\bm k}_+-{\bm q}',{\bm k}_-)
\nonumber\\
&& \times
{\cal D}_{\mu\mu''}({\bm k}'_-,{\bm k}'_+-{\bm q}')
{\cal D}_{\mu''\mu'}({\bm k}'_+-{\bm q}',{\bm k}'_+)
\nonumber\\
&& \times
G^{({\rm R},\sigma'')}_{\mu'}({\bm k}'_+,\varepsilon'+\omega) G^{({\rm A},\sigma'')}_{\mu}({\bm k}'_-,\varepsilon')
\nonumber\\
&& \times
\Lambda^{(\sigma''\sigma')}_{\mu'\mu,\beta} ({\bm k}'_+, \varepsilon'_++\omega, {\bm k}'_-,\varepsilon'_-)
~,
\end{eqnarray}
and finally
\begin{eqnarray} \label{eq:R_Lambda_3_together}
&&
\Lambda^{(3,\sigma\sigma')}_{\lambda'\lambda,\beta} ({\bm k}_+, \varepsilon_++\omega,{\bm k}_-,\varepsilon_-) =
-4 N_{\rm v} \sum_{{\bm k}',{\bm q}',\sigma''} \sum_{\mu,\mu'} \sum_{\lambda'',\mu''} 
\nonumber\\
&& \times
\int \frac{d\varepsilon'}{2\pi i} \int \frac{d\omega'}{2\pi i} 
W({\bm q}',\omega'_+) W({\bm q}',\omega'_--\omega)
\nonumber\\
&& \times
\big[n_{\rm F}(\varepsilon') + n_{\rm B}(\varepsilon' + \varepsilon)\big]
\big[n_{\rm F}(\omega'+\varepsilon) - n_{\rm F}(\omega'-\varepsilon'-\omega)\big]
\nonumber\\
&& \times
\Im m\big[ G^{({\rm R},\sigma)}_{\lambda''}({\bm k}_- +{\bm q}',\varepsilon+\omega') \big]
\nonumber\\
&& \times
\Im m\big[ G^{({\rm R},\sigma'')}_{\mu''}({\bm k}'_+ -{\bm q}',\varepsilon' +\omega- \omega') \big]
\nonumber\\
&& \times
{\cal D}_{\lambda\lambda''}({\bm k}_-,{\bm k}_-+{\bm q}')
{\cal D}_{\lambda''\lambda'}({\bm k}_-+{\bm q}',{\bm k}_+)
\nonumber\\
&& \times
{\cal D}_{\mu\mu''}({\bm k}'_-,{\bm k}'_+-{\bm q}')
{\cal D}_{\mu''\mu'}({\bm k}'_+-{\bm q}',{\bm k}'_+)
\nonumber\\
&& \times
G^{({\rm R},\sigma'')}_{\mu'} ({\bm k}'_+,\varepsilon'+\omega)
G^{({\rm A},\sigma'')}_{\mu} ({\bm k}'_-,\varepsilon')
\nonumber\\
&& \times
\Lambda^{(\sigma''\sigma')}_{\mu'\mu,\beta} ({\bm k}'_+, \varepsilon_++\omega,{\bm k}'_-,\varepsilon'_-)
~.
\end{eqnarray}
Equations~(\ref{eq:R_chi_jj_final_omega_finite})-(\ref{eq:R_Lambda_3_together}), together with the inverse quasiparticle lifetime defined in Eq.~(\ref{eq:inverse_lifetime_3}), constitute a closed set of equations that can be used to determine the retarded spin-resolved current-current response function in the limit $v_{\rm F} q\ll \omega, \tau_{\rm ee}^{-1} \ll \varepsilon_{\rm F}$. This calculation is performed in the next section.

\section{The charge and spin conductivities of graphene in the Fermi-liquid regime}
\label{sect:FL_conductivity}
In this section we present the derivation of the charge and spin conductivities of graphene. We start from Eqs.~(\ref{eq:R_chi_jj_final_omega_finite})-(\ref{eq:R_Lambda_3_together}), where we set $q=0$ and we take the limit $\omega \to 0$. To ${\cal O}(\omega)$ Eq.~(\ref{eq:R_chi_jj_final_omega_finite}) becomes
\begin{eqnarray} \label{eq:S_R_chi_jj_final_omega0}
&& \!\!\!\!\!\!\!\!
\chi_{j_{\alpha}^{(\sigma)} j_{\beta}^{(\sigma')}} ({\bm q} = {\bm 0}, \omega) = \omega N_{\rm v} \sum_{{\bm k}, \lambda, \lambda'} \int \frac{d\varepsilon}{2\pi i} \left(-\frac{\partial n_{\rm F} (\varepsilon)}{\partial \varepsilon} \right)
\nonumber\\
&\times&
G^{({\rm A},\sigma)}_{\lambda}({\bm k}, \varepsilon) \Lambda^{(0,\sigma)}_{\lambda\lambda',\alpha}({\bm k},{\bm k}) G^{({\rm R},\sigma)}_{\lambda'}({\bm k}, \varepsilon+\omega)
\nonumber\\
&\times&
\Lambda^{(\sigma\sigma')}_{\lambda'\lambda,\beta} ({\bm k}, \varepsilon_++\omega,{\bm k},\varepsilon_-)
~.
\end{eqnarray}
The limit $\omega \to 0$ is understood in this equation.
We observe that the function $-\partial n_{\rm F} (\varepsilon)/\partial \varepsilon$ is peaked around $\varepsilon=0$ and tends to a $\delta$-function in the low-temperature limit $\varepsilon_{\rm F} \tau_{\rm ee} \gg 1$. We thus evaluate all the other functions on the right-hand side of Eq.~(\ref{eq:S_R_chi_jj_final_omega0}) at $\varepsilon = 0$, with the exception of $\Lambda^{(\sigma\sigma')}_{\lambda'\lambda,\beta} ({\bm k}, \varepsilon_+,{\bm k},\varepsilon_-)$, which requires further care. As it will become clear in what follows, the latter contains Fermi and Bose factors which depend on $\varepsilon$ and that combine with $\partial n_{\rm F} (\varepsilon)/\partial \varepsilon$ in Eq.~(\ref{eq:S_R_chi_jj_final_omega0}) to yield the correct transport times. Missing this step would lead to a non-cancellation between the self-energy and vertex corrections in the charge channel.
In the limit $\omega \to 0$ we also approximate
\begin{equation} \label{eq:S_R_GR_GA_product}
G^{({\rm A},\sigma)}_\lambda({\bm k}, 0) G^{({\rm R},\sigma)}_{\lambda'}({\bm k}, \omega) \simeq
-\frac{2 i \delta_{\lambda\lambda'} }{\omega + i/\tau_{\rm ee}}
\Im m\big[ G^{({\rm R},\sigma)}_{\lambda}({\bm k}, 0) \big]
~.
\end{equation}
In so doing  we neglect the incoherent part of  the Green's function, i.e., the part of $G$ that is not included in the quasiparticle-pole approximation.  Herein lies our Fermi liquid approximation.
At low temperature, $\Im m\big[ G^{({\rm R})}_{\lambda}({\bm k}, 0) \big]$ is a Lorentzian strongly peaked around $\varepsilon_{{\bm k},\lambda} = \varepsilon_{\rm F}$ and with a width proportional to the quasiparticle decay rate. This implies that $k\sim k_{\rm F}$ and $\lambda=\lambda'=+$. For $\varepsilon_{\rm F} \tau_{\rm ee} \gg 1$ the transport is dominated by states that lie in a thin shell of thickness $\simeq k_{\rm B} T$ around the Fermi energy~\cite{Giuliani_and_Vignale}.

At the same level of approximation, the dressed vertex function satisfies the following Bethe-Salpeter equation (see also App.~\ref{app:BS_spin}):
\begin{eqnarray} \label{eq:S_Bethe_Salpeter_final}
&&
\Lambda_{++,\beta}^{(\sigma\sigma')} ({\bm k}, \varepsilon_++\omega, {\bm k}, \varepsilon_-) = \delta_{\sigma\sigma'}\Lambda^{(0,\sigma)}_{++,\beta} ({\bm k},{\bm k})
-\frac{8 i N_{\rm v}}{\omega + i/\tau_{\rm ee}}
\nonumber\\
&& \times
\sum_{{\bm k}',{\bm q}',\sigma''}|W({\bm q}',0)|^2
\int \frac{d\varepsilon'}{2\pi i} \int \frac{d\omega'}{2\pi i}
\big[n_{\rm F}(\varepsilon') + n_{\rm B}(\varepsilon'-\varepsilon)\big]
\nonumber\\
&& \times
\big[n_{\rm F} (\omega'+\varepsilon') - n_{\rm F} (\omega'+\varepsilon)\big] 
\Im m \big[ G^{({\rm R},\sigma'')}_{+}({\bm k}',0) \big]
\nonumber\\
&& \times
\Im m \big[ G^{({\rm R},\sigma)}_{+}({\bm k}-{\bm q}',0) \big]
\Im m\big[G^{({\rm R},\sigma'')}_{+}({\bm q}'-{\bm k}', 0) \big]
\nonumber\\
&& \times
{\cal D}_{++}({\bm k},{\bm k}-{\bm q}')
{\cal D}_{++}({\bm k}-{\bm q}',{\bm k})
{\cal D}_{++}({\bm k}',{\bm k}'-{\bm q}')
\nonumber\\
&& \times
{\cal D}_{++}({\bm k}'-{\bm q}',{\bm k}') 
\Big[
\Lambda^{(\sigma\sigma')}_{++,\beta} ({\bm k}-{\bm q}', \varepsilon'_++\omega, {\bm k}-{\bm q}',\varepsilon'_-)
\nonumber\\
&&
+ \Lambda^{(\sigma''\sigma')}_{++,\beta} ({\bm k}', \varepsilon'_++\omega, {\bm k}',\varepsilon'_-)
\nonumber\\
&& 
-\Lambda^{(\sigma''\sigma')}_{++,\beta} ({\bm k}'-{\bm q}', \varepsilon'_++\omega,{\bm k}'-{\bm q}',\varepsilon'_-)
\Big]
~.
\nonumber\\
\end{eqnarray}
Here the limits $k=k_{\rm F}$ and $\omega\to 0$ are understood. In deriving Eq.~(\ref{eq:S_Bethe_Salpeter_final}) we used that in the low-temperature limit also the momenta ${\bm k}'$, ${\bm k}-{\bm q}'$ and ${\bm k}'-{\bm q}'$ are all pinned at the Fermi surface, {\it i.e.} $|{\bm k}'|=|{\bm k}-{\bm q}'|=|{\bm k}'-{\bm q}'|=k_{\rm F}$, and that the corresponding quasiparticles live in the conduction band (recall that the system is n-doped). Thus Eq.~(\ref{eq:S_Bethe_Salpeter_final}) is a closed self-consistent equation for the dressed vertex with all momentum- and energy-arguments pinned at the Fermi surface.

Equation~(\ref{eq:S_Bethe_Salpeter_final}) can be further simplified by noting that its solution must be introduced into Eq.~(\ref{eq:S_R_chi_jj_final_omega0}). We can thus immediately carry out the angular integration over $\varepsilon$ with the weighting function $\partial n_{\rm F}(\varepsilon)/\partial\varepsilon$, and we get
\begin{eqnarray} \label{eq:S_Bethe_Salpeter_final_2}
&&
\Lambda_{++,\beta}^{(\sigma\sigma')} ({\bm k}, \omega^+, {\bm k}, 0^-) = \delta_{\sigma\sigma'}\Lambda^{(0,\sigma)}_{++,\beta} ({\bm k},{\bm k})
- \frac{4 i N_{\rm v} (k_{\rm B} T)^2}{3(\omega + i/\tau_{\rm ee})}
\nonumber\\
&& \times
\sum_{{\bm k}',{\bm q}',\sigma''}|W({\bm q}',0)|^2
\Im m \big[ G^{({\rm R},\sigma'')}_{+}({\bm k}',0) \big]
\nonumber\\
&& \times
\Im m \big[ G^{({\rm R},\sigma)}_{+}({\bm k}-{\bm q}',0) \big]
\Im m\big[G^{({\rm R},\sigma'')}_{+}({\bm q}'-{\bm k}', 0) \big]
\nonumber\\
&& \times
{\cal D}_{++}({\bm k},{\bm k}-{\bm q}')
{\cal D}_{++}({\bm k}-{\bm q}',{\bm k})
{\cal D}_{++}({\bm k}',{\bm k}'-{\bm q}')
\nonumber\\
&& \times
{\cal D}_{++}({\bm k}'-{\bm q}',{\bm k}') 
\Big[
\Lambda^{(\sigma\sigma')}_{++,\beta} ({\bm k}-{\bm q}', \omega^+, {\bm k}-{\bm q}',0^-)
\nonumber\\
&&
+ \Lambda^{(\sigma''\sigma')}_{++,\beta} ({\bm k}', \omega^+, {\bm k}',0^-)
\nonumber\\
&&
-\Lambda^{(\sigma''\sigma')}_{++,\beta} ({\bm k}'-{\bm q}', \omega^+,{\bm k}'-{\bm q}',0^-)
\Big]
~.
\end{eqnarray}
To get this equation we used the results of Eqs.~(\ref{eq:nB_nF_approx})-(\ref{eq:energy_approx}).

We now solve Eq.~(\ref{eq:S_Bethe_Salpeter_final_2}) with standard methods~\cite{Bruus_and_Flensberg}. We first reduce it to an algebraic equation with the following {\it Ansatz}:
\begin{equation} \label{eq:S_Lambda_ansatz}
\Lambda^{(\sigma\sigma')}_{++,\beta} ({\bm k}, \omega^+, {\bm k},0^-) = \gamma_{\sigma\sigma'}(\omega) \Lambda^{(0,\sigma')}_{++,\beta} ({\bm k},{\bm k}) 
~,
\end{equation}
and we then solve it for $\gamma_{\sigma\sigma'}(\omega)$. Note that at the Fermi surface $\Lambda^{(0,\sigma')}_{++,\beta} ({\bm k},{\bm k}) = v_{\rm F} {\hat {\bm k}}_\beta$. A further simplification comes from the fact that the Green's functions on the right-hand side of Eq.~(\ref{eq:S_Bethe_Salpeter_final_2}) are independent of spin. We suppress their spin dependence in what follows. 

Let us first consider the Bethe-Salpeter Eq.~(\ref{eq:S_Bethe_Salpeter_final_2}) in the charge channel. To obtain the dressed charge vertex, after having introduced the {\it Ansatz}~(\ref{eq:S_Lambda_ansatz}) into Eq.~(\ref{eq:S_Bethe_Salpeter_final_2}), we sum over the spin index $\sigma$. After some algebra we obtain
\begin{eqnarray} \label{eq:S_Bethe_Salpeter_final_plus}
&&
\big[\gamma_{+\sigma'}(\omega)+\gamma_{-\sigma'}(\omega)\big] \Lambda^{(0,\sigma')}_{++,\beta} ({\bm k},{\bm k})  = \Lambda^{(0,\sigma')}_{++,\beta} ({\bm k},{\bm k})
\nonumber\\
&&
-\frac{8 i N_{\rm v} (k_{\rm B} T)^2}{3(\omega + i/\tau_{\rm ee})} \big[\gamma_{+\sigma'}(\omega)+\gamma_{-\sigma'}(\omega)\big]
\sum_{{\bm k}',{\bm q}'}|W({\bm q}',0)|^2
\nonumber\\
&& \times
\Im m \big[ G^{({\rm R})}_{+}({\bm k}',0) \big]
\Im m \big[ G^{({\rm R})}_{+}({\bm k}-{\bm q}',0) \big]
\nonumber\\
&& \times
\Im m\big[G^{({\rm R})}_{+}({\bm q}'-{\bm k}', 0) \big]
{\cal D}_{++}({\bm k},{\bm k}-{\bm q}')
\nonumber\\
&& \times
{\cal D}_{++}({\bm k}-{\bm q}',{\bm k})
{\cal D}_{++}({\bm k}',{\bm k}'-{\bm q}')
{\cal D}_{++}({\bm k}'-{\bm q}',{\bm k}') 
\nonumber\\
&& \times
\Big[
\Lambda^{(0,\sigma')}_{++,\beta} ({\bm k}-{\bm q}', {\bm k}-{\bm q}')
+ \Lambda^{(0,\sigma')}_{++,\beta} ({\bm k}', {\bm k}')
\nonumber\\
&&
-\Lambda^{(0,\sigma')}_{++,\beta} ({\bm k}'-{\bm q}',{\bm k}'-{\bm q}')
\Big]
~.
\end{eqnarray}
Let us now consider Eq.~(\ref{eq:S_Bethe_Salpeter_final_2}) in the spin channel. In this case we first multiply by $\sigma$ and then we sum over the spin index $\sigma$. We get
\begin{eqnarray} \label{eq:S_Bethe_Salpeter_final_minus}
&&
\big[\gamma_{+\sigma'}(\omega) - \gamma_{-\sigma'}(\omega)\big] \Lambda^{(0,\sigma')}_{++,\beta} ({\bm k},{\bm k})  = \sigma'\Lambda^{(0,\sigma')}_{++,\beta} ({\bm k},{\bm k})
\nonumber\\
&&
- \frac{8 i N_{\rm v} (k_{\rm B} T)^2}{3(\omega + i/\tau_{\rm ee})} \big[\gamma_{+\sigma'}(\omega)-\gamma_{-\sigma'}(\omega)\big]
\sum_{{\bm k}',{\bm q}'}|W({\bm q}',0)|^2
\nonumber\\
&& \times
\Im m \big[ G^{({\rm R})}_{+}({\bm k}',0) \big]
\Im m \big[ G^{({\rm R})}_{+}({\bm k}-{\bm q}',0) \big]
\nonumber\\
&& \times
\Im m\big[G^{({\rm R})}_{+}({\bm q}'-{\bm k}', 0) \big]
{\cal D}_{++}({\bm k},{\bm k}-{\bm q}')
\nonumber\\
&& \times
{\cal D}_{++}({\bm k}-{\bm q}',{\bm k})
{\cal D}_{++}({\bm k}',{\bm k}'-{\bm q}')
{\cal D}_{++}({\bm k}'-{\bm q}',{\bm k}') 
\nonumber\\
&& \times
\Lambda^{(0,\sigma')}_{++,\beta} ({\bm k}-{\bm q}', {\bm k}-{\bm q}')
~.
\end{eqnarray}
Eq.~(\ref{eq:S_Bethe_Salpeter_final_plus}) and~(\ref{eq:S_Bethe_Salpeter_final_minus}) control the vertex renormalization in the charge and spin channels, respectively. Note that the former depends only on the variable $\gamma_{+\sigma'}(\omega) + \gamma_{-\sigma'}(\omega)$, while the latter depends only on $\gamma_{+\sigma'}(\omega) - \gamma_{-\sigma'}(\omega)$, and that they are thus independent of each other. The spin and charge channels are thus decoupled in an unpolarized system.

\begin{figure}[t]
\begin{center}
\begin{tabular}{c}
\includegraphics[width=0.99\columnwidth]{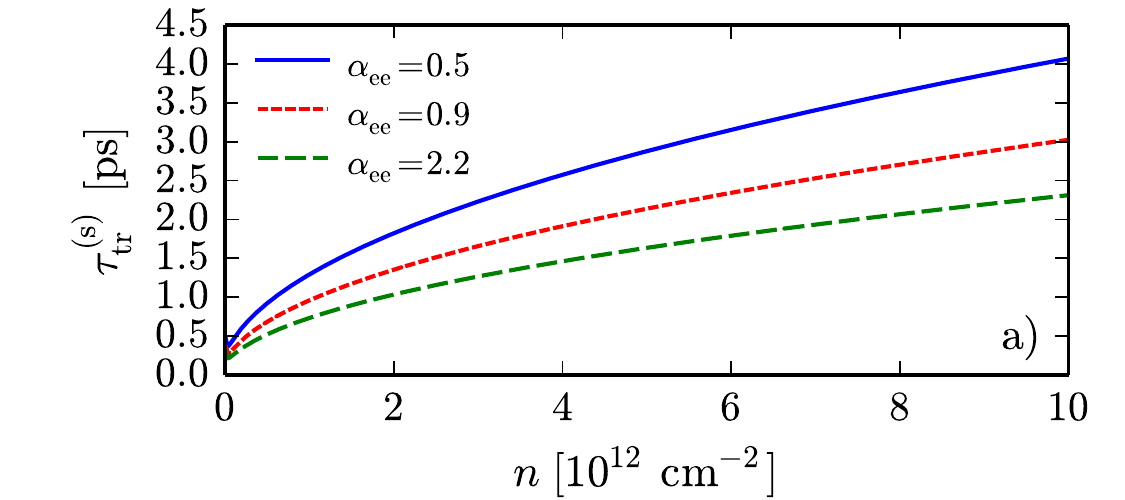}
\\
\includegraphics[width=0.99\columnwidth]{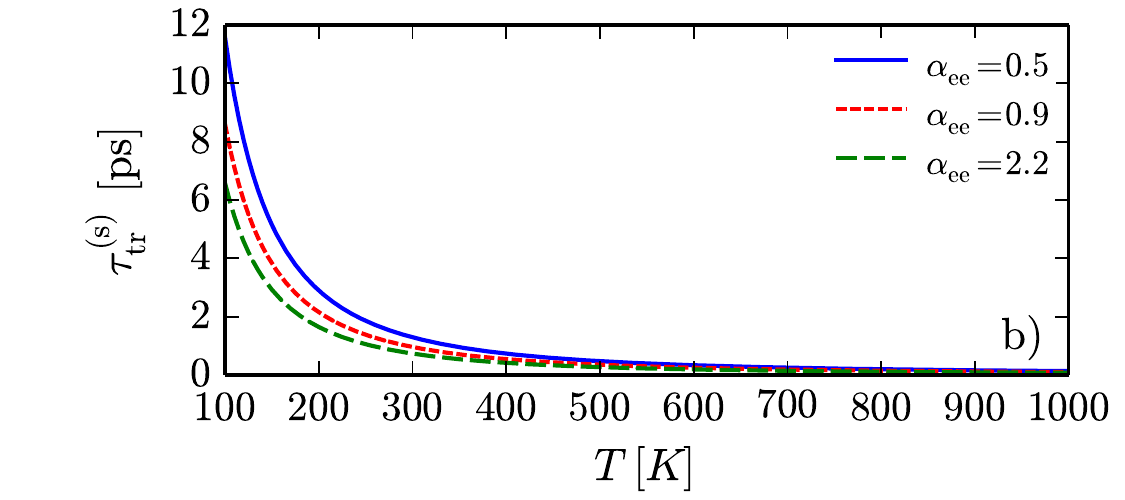}
\end{tabular}
\end{center}
\caption{
(Color online) Panel a) the spin transport time of massless Dirac fermions $\tau_{\rm tr}^{({\rm s})}$, as defined in Eq.~(\ref{eq:S_tau_tr}), in units of picoseconds and plotted as a function of the density $n$ in units of $10^{12}~cm^{-2}$ for three values of the dimensionless coupling constant $\alpha_{\rm ee}$. In this plot we fixed the temperature $T = 300~{\rm K}$. Panel b) same as in panel a) but shown as a function of temperature (in units of ${\rm K}$) for a fixed excess carrier density $n=10^{12}~{\rm cm}^{-2}$.
\label{fig:seven}}
\end{figure}

We are now in the position to solve Eqs.~(\ref{eq:S_Bethe_Salpeter_final_plus})-(\ref{eq:S_Bethe_Salpeter_final_minus}). We project them along the direction of $\Lambda^{(0,\sigma')}_{++,\beta} ({\bm k},{\bm k}) = v_{\rm F} {\hat {\bm k}}_\beta$ and, after some straightforward algebraic manipulations, we find
\begin{equation} \label{eq:S_charge_vertex}
\gamma_{+\sigma'}(\omega) + \gamma_{-\sigma'}(\omega) = \frac{\omega+i/\tau_{\rm ee}}{\omega+i\eta}
~,
\end{equation}
and
\begin{equation} \label{eq:S_spin_vertex}
\gamma_{+\sigma'}(\omega) - \gamma_{-\sigma'}(\omega) = \sigma'\frac{\omega+i/\tau_{\rm ee}}{\omega+i/\tau^{(s)}_{\rm tr}}
~.
\end{equation}
We recall that the quasiparticle lifetime at the Fermi surface, $\tau_{\rm ee}$, is defined in Eq.~(\ref{eq:inverse_lifetime_3}) and it is explicitly calculated in Fig.~\ref{fig:four}. In Eq.~(\ref{eq:S_spin_vertex}) we defined
\begin{eqnarray} \label{eq:S_tau_tr}
\frac{1}{\tau^{({\rm s})}_{\rm tr}}  &=& 
- \frac{8}{3}  N_{\rm v} (k_{\rm B} T)^2
\sum_{{\bm k}',{\bm q}'}|W({\bm q}',0)|^2
\Im m \big[ G^{({\rm R})}_{+}({\bm k}-{\bm q}',0) \big]
\nonumber\\
&\times&
\Im m \big[ G^{({\rm R})}_{+}({\bm k}',0) \big]
\Im m\big[G^{({\rm R})}_{+}({\bm q}'-{\bm k}', 0) \big]
\nonumber\\
&\times&
{\cal D}_{++}({\bm k},{\bm k}-{\bm q}')
{\cal D}_{++}({\bm k}-{\bm q}',{\bm k})
\nonumber\\
&\times&
{\cal D}_{++}({\bm k}',{\bm k}'-{\bm q}')
{\cal D}_{++}({\bm k}'-{\bm q}',{\bm k}') 
\nonumber\\
&\times&
\big[1 - \cos(\varphi_{\bm k} - \varphi_{{\bm k}-{\bm q}'})\big]
~.
\end{eqnarray}
The calculation of Eq.~(\ref{eq:S_tau_tr}) is performed in App.~\ref{app:spin_transport_time}. Numerical results for $\tau_{\rm tr}^{({\rm s})}$ are shown in Fig.~\ref{fig:seven}.

We substitute Eqs.~(\ref{eq:S_Lambda_ansatz}), with $\gamma_{\sigma\sigma'}(\omega)$ defined in Eqs.~(\ref{eq:S_charge_vertex}) and~(\ref{eq:S_spin_vertex}), back into the definition of the spin-resolved current-current response function of Eq.~(\ref{eq:S_R_chi_jj_final_omega0}). From this we then compute the charge and spin conductivities, according to the definitions given in Eqs.~(\ref{eq:sigma_c_def}) and~(\ref{eq:sigma_s_def}). After some lengthy but straightforward algebra we get (we restore $\hbar$ in the following expressions)
\begin{eqnarray} \label{eq:R_S_chi_jj_final}
\sigma_{\alpha\beta}^{({\rm c})} (\omega) &=& N_{\rm v} \frac{e^2}{\hbar} \sum_{{\bm k}, \lambda, \lambda'} \int \frac{d\varepsilon}{2\pi}
\left(-\frac{\partial n_{\rm F} (\varepsilon)}{\partial \varepsilon}\right)
G^{({\rm A})}_\lambda({\bm k}, \varepsilon) 
\nonumber\\
&\times&
\Lambda^{(0)}_{\lambda,\lambda',\alpha}({\bm k},{\bm k})
G^{({\rm R})}_{\lambda'}({\bm k}, \varepsilon+\omega) \Lambda^{(0)}_{\lambda'\lambda,\beta} ({\bm k}, {\bm k})
\nonumber\\
&\times&
\sum_{\sigma,\sigma'=\pm} \gamma_{\sigma\sigma'}(\omega)
\nonumber\\
&=&
\delta_{\alpha\beta} \frac{e^2 {\cal D}^{(0)}_{\rm c}}{-i\omega+\eta}
~.
\end{eqnarray}
In Eq.~(\ref{eq:R_S_chi_jj_final}) we defined the non-interacting charge Drude weight ${\cal D}^{(0)}_{\rm c} = N_{\rm v} \varepsilon_{\rm F}/(2\pi\hbar^2)$. Note that in this equation the quasiparticle lifetime disappears as a consequence of a cancellation that occurs between the product of the two Green's functions, approximated as in Eq.~(\ref{eq:S_R_GR_GA_product}), and the vertex correction~(\ref{eq:S_charge_vertex}). 

Equation~(\ref{eq:R_S_chi_jj_final}) shows that the real part of the DC charge conductivity of interacting graphene is infinite. Moreover, the weight of the low-frequency Drude peak coincides with the non-interacting one. As discussed in Sect.~\ref{sect:self_energy}, our theory does not capture the renormalization of the Drude weight due to electron-electron interactions~\cite{kotov_rmp_2012,abedinpour_prb_2011,Levitov_prb_2013}, since we have neglected the contribution of the real part of the one-body self-energy. Such contribution can be  taken into account by replacing
\begin{equation} \label{eq:charge_drude_weight}
{\cal D}^{(0)}_{\rm c} \to {\cal D}_{\rm c} \equiv \frac{N_{\rm v}}{2\pi \hbar^2} (1 + F_1^{\rm s}) v_{\rm F}^\star k_{\rm F}
~.
\end{equation}
Note that electron-electron interactions affect the Drude weight in a two-fold way. On one hand the bare Fermi velocity is replaced by its interacting counterpart $v_{\rm F}^\star$, which embeds the self-energy corrections at the single-particle level. On the other hand, electron-electron interactions provide also ``vertex corrections'' to two-body properties. In the case of the charge conductivity these are encoded in the Landau parameter~\cite{Giuliani_and_Vignale} $F_1^{\rm s}$. Both the self-energy and vertex corrections to the charge Drude weight have been calculated to the first order in the strength of electron-electron interactions in Ref.~\onlinecite{abedinpour_prb_2011}.

Our calculations show that {\it doped} graphene at low temperature  behaves as an {\it effectively} Galilean-invariant system.  Indeed, if graphene is doped and the temperature is sufficiently low, the velocity and momentum of the current-carrying states, {\it i.e.} the quasiparticles at the Fermi surface, are linearly related via the Fermi wavevector $k_{\rm F}$. This in turn implies that the current carried by such states is proportional to their total momentum. 
As in a parabolic-band electron gas, electron-electron interactions conserve the total momentum and are thus inefficient in relaxing the current carried by quasiparticles at the Fermi surface.
We emphasize that this conclusion is reached for  {\it doped} graphene.    As the doping level decreases at a given temperature the momentum-velocity relation can no longer be linearized and the behavior of the conductivities must be reconsidered.  This is in all likelihood the reason why pristine (undoped)  graphene exhibits a finite charge conductivity, as opposed to the infinite conductivity discussed here.

Following  steps analogous to those taken in the derivation of Eq.~(\ref{eq:R_S_chi_jj_final}), we get the spin conductivity
\begin{eqnarray} \label{eq:S_conductivity}
\sigma_{\alpha\beta}^{({\rm s})} (\omega) &=& N_{\rm v} \hbar \sum_{{\bm k}, \lambda, \lambda'} \int \frac{d\varepsilon}{2\pi}
\left(-\frac{\partial n_{\rm F} (\varepsilon)}{\partial \varepsilon}\right)
G^{({\rm A})}_\lambda({\bm k}, \varepsilon) 
\nonumber\\
&\times&
\Lambda^{(0,\alpha)}_{\lambda,\lambda'}({\bm k},{\bm k})
G^{({\rm R})}_{\lambda'}({\bm k}, \varepsilon) \Lambda^{(0,\beta)}_{\lambda'\lambda} ({\bm k}, {\bm k})
\nonumber\\
&\times&
\sum_{\sigma,\sigma'=\pm} \sigma \sigma'\gamma_{\sigma\sigma'}(k_{\rm F})
\nonumber\\
&=&
\delta_{\alpha\beta} \frac{\hbar^2 {\cal D}^{(0)}_{\rm s}}{-i\omega+1/\tau^{(\rm s)}_{\rm tr}}
~.
\end{eqnarray}
Where the non-interacting spin-Drude weight ${\cal D}^{(0)}_{\rm s}$ coincides with ${\cal D}^{(0)}_{\rm c}$.
Again, our result misses the renormalization of the spin-Drude weight, which is calculated in Ref.~\onlinecite{Qian_prl_2004} for a parabolic-band electron gas and in App.~\ref{eq:spin_velocity} for graphene and reads
\begin{equation} \label{eq:spin_drude_weight}
{\cal D}_{\rm s} \equiv \frac{1+F_1^{\rm a}}{1 + F_1^{\rm s}} {\cal D}_{\rm c}
~.
\end{equation}
As it happens in a 2DEG, this renormalization is expected to be small also in graphene, for typical carrier densities and in a broad range of values of the coupling constant $\alpha_{\rm ee}$. The finiteness of the relaxation rate is thus by far the largest effect of electron-electron interactions on the spin transport in graphene. The spin conductivity becomes finite when particle-particle interactions are turned on, even in a perfectly clean and infinite system, and scales as $\sim T^{-2}$ in the DC limit ($\omega \to 0$).
Finally, we can calculate the spin-drag transresistivity~\cite{DAmico_prb_2003} $\rho_{\uparrow\downarrow}$ in the limit $k_{\rm B} T \ll \varepsilon_{\rm F}$, according to the definition $\rho_{\uparrow\downarrow} = -\big[\tau^{({\rm s})}_{\rm tr} {\cal D}_{\rm s}/e^2]^{-1}$. Neglecting the Fermi-liquid renormalization of the Drude weight, For $k_{\rm B}T/\varepsilon_{\rm F} = 0.1$ and for a coupling constant $\alpha_{\rm ee}=2.2$ we get $|\rho_{\uparrow\downarrow}| \sim 54~\Omega$. As a comparison~\cite{DAmico_prb_2003}, in a 2DEG with $r_s = \alpha_{\rm ee}$ and same ratio $k_{\rm B}T/\varepsilon_{\rm F}$, $|\rho_{\uparrow\downarrow}| \sim 200~\Omega$.

\section{Summary and conclusions}
\label{sect:summary}
In this Paper we have calculated two fundamental transport properties of doped graphene, the charge and spin conductivities, in the ``hydrodynamic'' regime of the electron liquid~\cite{Andreev_prl_2011} --- see Eqs.~(\ref{eq:R_S_chi_jj_final}) and~(\ref{eq:S_conductivity}). In this regime the rate of electron-electron collisions is extremely high, and the other scattering mechanisms, like electron-phonon or electron-impurity interactions, are negligible~\cite{Fritz_prb_2008_1,Fritz_prb_2008_2}. Electron-electron collisions bring the system to a state of local quasi-equilibrium, which is described by a set of smoothly varying (in space and time) functions, {\it i.e.} the density, the velocity field and the local temperature~\cite{Landau_6}. Such intriguing regime is rarely relevant in solids~\cite{Andreev_prl_2011}, since momentum-non-conserving processes dominate the charge transport. Very low temperatures and clean samples are needed to expose it in experiments. 

Graphene stands out, among known materials, as the ultimate hydrodynamic material. Ultra-high-quality samples of graphene deposited on hexagonal Boron Nitride, which can be produced by  standard techniques, are indeed virtually free of long- and short-range impurities and of charge inhomogeneities.~\cite{Mayorov_nanolett_2012} This, combined with the linear band dispersion and the lattice stiffness,~\cite{Balandin_NatMat_2011} yields a fairly large temperature window in which the hydrodynamic regime can be realized. Indeed, due to the high lattice stiffness,  scattering with in-plane acoustic phonons becomes relevant only at temperatures of the order of the room temperature~\cite{footnote_time_scales}.

Hydrodynamics is usually defined for Galilean-invariant systems~\cite{Landau_6}, in which the conservation of momentum leads to a diverging charge conductivity even in the presence of electron-electron interactions. However, at low energies graphene is well-described by a Dirac Hamitonian, which is not Galilean invariant. This reflects in the fact that a homogenous current is not proportional to the total momentum, and can in principle be relaxed by electron-electron interactions, thus invalidating the whole hydrodynamic picture. 
Our calculations show that, while the spin conductivity is finite and limited by electron-electron interactions, the {\it homogeneous} charge conductivity is infinite as for an interacting 2DEG. The latter result stems from the fact that, at low temperature, doped graphene behaves as an effectively Galilean-invariant system, and paves the way to the description of charge transport in graphene in terms of Navier-Stokes equations. The relaxation of {\it inhomogeneous} current distributions is controlled by the viscosities of the electron liquid in graphene, which will be given in a forthcoming publication.

As the doping level decreases and graphene approaches the {\it undoped} regime, the momentum-velocity relation can no longer be linearized. Thus, not only the Drude weight gets renormalized, but also the charge transport time becomes finite because of electron-electron interactions~\cite{schutt_prb_2011}. This in turn implies that, in this regime, the evolution of the charge current is described by a ``generalized'' Navier-Stokes equation which contains a relaxation term.
While this Manuscript was being prepared, we became aware that a hydrodynamic theory of undoped graphene was worked out in Ref.~\onlinecite{Narozhny_arxiv_2014}. To describe the transport in this system in the collision-dominated limit, it is necessary to consider {\it simultaneously} the equations of motion of three macroscopic currents, {\it i.e.} the charge, energy and ``quasiparticle-imbalance'' currents~\cite{Narozhny_arxiv_2014}. Note that the energy current, being proportional to the momentum in systems with a linear band dispersion, is always strictly conserved by electron-electron interactions and follows a Navier-Stokes equation. Conversely, the equation of motion of the charge current contains a finite relaxation rate due to electron-electron interactions. In the doped regime the three currents coincide~\cite{Narozhny_arxiv_2014} and the relaxation rate of the charge current due to electron-electron interactions vanishes. Thus, the transport can be described by taking into account only the charge current, whose evolution is now described by the same Navier-Stokes equation satisfied by the energy current.

\section{Acknowledgments}
This work was supported in part by DOE grant DE-FG02-05ER46203 and by a Research Board Grant at the University of Missouri.

\appendix

\section{The calculation of the quasiparticle lifetime}
\label{app:quasiparticle_lifetime}
Before deriving the quasiparticle lifetime, let us recall Eqs.~(\ref{eq:GW})-(\ref{eq:chi_nn}) which define the GW self-energy. They read
\begin{eqnarray} \label{eq:app_GW}
\Sigma^{(\sigma)}_\lambda({\bm k},i\varepsilon_n) &=& -k_{\rm B} T \sum_{{\bm k}',\lambda'} \sum_{\varepsilon_{n'}} W({\bm k}'-{\bm k},i\varepsilon_{n'} - i\varepsilon_n)
\nonumber\\
&\times&
G_{\lambda'}^{(\sigma)}({\bm k}',i\varepsilon_{n'}) {\cal D}_{\lambda\lambda'}({\bm k},{\bm k}') {\cal D}_{\lambda'\lambda}({\bm k}',{\bm k})
~,
\nonumber\\
\end{eqnarray}
and
\begin{equation} \label{eq:app_W_RPA}
W({\bm q},i\Omega_{m}) = \frac{v_{\bm q}}{1-v_{\bm q} \chi_{nn}({\bm q},i\Omega_m)}
~,
\end{equation}
and finally
\begin{eqnarray} \label{eq:app_chi_nn}
\chi_{nn}({\bm q},i\omega_m) &=& N_{\rm v} k_{\rm B} T \sum_{{\bm q}',\varepsilon_n,\sigma'} \sum_{\lambda'',\mu''} G_{\lambda''}^{(\sigma')}({\bm q}',i\varepsilon_n)
\nonumber\\
&\times&
G_{\mu''}^{(\sigma')}({\bm q'}+{\bm q}, i\varepsilon_n+i\omega_m)
\nonumber\\
&\times&
{\cal D}_{\lambda''\mu''}({\bm q}',{\bm q}'+{\bm q}) {\cal D}_{\mu''\lambda''}({\bm q}'+{\bm q},{\bm q}') 
~.
\nonumber\\
\end{eqnarray}

We first consider Eq.~(\ref{eq:app_GW}), and we analytically continue it to real frequencies. We thus define $f_\Sigma(i\varepsilon_{n'} - i\varepsilon_n, i\varepsilon_{n'})$ such that
\begin{eqnarray}
\Sigma^{(\sigma)}({\bm k},i\varepsilon_n) &\equiv& -k_{\rm B} T \sum_{\varepsilon_{n'}} f_\Sigma(i\varepsilon_{n'} - i\varepsilon_n, i\varepsilon_{n'})
\nonumber\\
&=&
\oint_{\cal C} \frac{dz}{2\pi i} n_{\rm F}(z) f_\Sigma(z - i\varepsilon_n, z)
~.
\end{eqnarray}
The contour ${\cal C}$ in the complex plane encircles all the poles of the Fermi function $n_{\rm F}(z) = \big[\exp(\beta z)+1\big]^{-1}$, and leaves outside the branch cuts of $f_\Sigma(z - i\varepsilon_n, z)$, which are parallel to the real axis and pass through $z=0, i \varepsilon_n$. Deforming the contour of integration to encircle the branch cuts of $f_\Sigma(z - i\varepsilon_n, z)$, and taking the limit $i\varepsilon_n \to \varepsilon + i\eta$ we obtain the retarded self-energy~\cite{Giuliani_and_Vignale}
\begin{eqnarray} \label{eq:app_sigma_retarded}
\Sigma^{(\sigma)}({\bm k},\varepsilon_+) &=&
\int_{-\infty}^{\infty} \frac{d\varepsilon'}{2\pi i} \Big\{ 
\big[n_{\rm F}(\varepsilon') + n_{\rm B}(\varepsilon'-\varepsilon)\big] 
\nonumber\\
&\times&
\big[f_\Sigma(\varepsilon'_- - \varepsilon, \varepsilon'_+) - f_\Sigma(\varepsilon'_+ - \varepsilon, \varepsilon'_+)\big]
\nonumber\\
&+&
n_{\rm F}(\varepsilon') \big[ f_\Sigma(\varepsilon'_- - \varepsilon, \varepsilon'_-) - f_\Sigma(\varepsilon'_+ - \varepsilon, \varepsilon'_+) \big]
\Big\}
~.
\nonumber\\
\end{eqnarray}
Here $\varepsilon_\pm = \varepsilon \pm i \eta$. The term in the last line of Eq.~(\ref{eq:app_sigma_retarded}) is purely real, since it is multiplied by the imaginary unit. We thus get
\begin{eqnarray} \label{eq:app_taum1}
\frac{1}{\tau_{\rm ee}} &=& -2\int_{-\infty}^{\infty} d\varepsilon \int_{-\infty}^{\infty} \frac{d\varepsilon'}{\pi} \frac{\partial n_{\rm F}(\varepsilon)}{\partial \varepsilon} \big[n_{\rm F}(\varepsilon') + n_{\rm B}(\varepsilon'-\varepsilon)\big]
\nonumber\\
&\times&
\sum_{{\bm k}',\lambda,\lambda'}  \Im m G^{(\sigma)}({\bm k}',\varepsilon'_+)
\Im m W({\bm k}-{\bm k}', \varepsilon'_+-\varepsilon) 
\nonumber\\
&\times&
{\cal D}_{\lambda\lambda'}({\bm k},{\bm k}') {\cal D}_{\lambda'\lambda}({\bm k}',{\bm k})
~.
\end{eqnarray}
Here we understand that $|{\bm k}| = k_{\rm F}$. The imaginary part of the screened interaction reads
\begin{equation} \label{eq:app_Im_W}
\Im m W({\bm q}, \omega_+) = |W({\bm q}, \omega_+)| \Im m \chi_{nn}({\bm q}, \omega_+)
~.
\end{equation}
The analytical continuation of $\chi_{nn}({\bm q}, i\omega_m)$ defined in Eq.~(\ref{eq:app_chi_nn}) can be performed in analogy with that of the self-energy. We thus define
\begin{eqnarray} 
\chi_{nn}({\bm q},i\omega_m) &=& k_{\rm B} T \sum_{\varepsilon_n} f_\chi(i\varepsilon_{n} + i\omega_m, i\varepsilon_{n})
\nonumber\\
&=&
-\oint_{{\cal C}'} \frac{dz}{2\pi i} n_{\rm F}(z) f_\chi(z + i\omega_m, z)
~.
\end{eqnarray}
Also the contour ${\cal C}'$ encircles only the poles of the Fermi function and leaves outside the branch cuts of $f_\chi(z + i\omega_n, z)$, which are parallel to the real axis and pass through $z=0, -i \omega_m$. Deforming the contour of integration to encircle the branch cuts of $f_\chi(z + i\omega_n, z)$, and taking the limit $i\omega_m \to \omega+i\eta$ we get
\begin{eqnarray} 
\chi_{nn}({\bm q},\omega_+) &=&  -\int_{-\infty}^{\infty} \frac{d\varepsilon''}{2\pi i} \Big\{ \big[ n_{\rm F}(\varepsilon''+\omega) - n_{\rm F}(\varepsilon'') \big]
\nonumber\\
&\times&
\big[ f(\varepsilon''_++\omega,\varepsilon''_-) - f(\varepsilon''_-+\omega,\varepsilon''_-) \big]
\nonumber\\
&+&
n_{\rm F}(\varepsilon'') \big[ f(\varepsilon''_++\omega,\varepsilon''_+) - f(\varepsilon''_-+\omega,\varepsilon''_-) \big]
\Big\}
~.
\nonumber\\
\end{eqnarray}
Again the last term gives no contribution to the imaginary part, which becomes
\begin{eqnarray} \label{eq:app_Im_chi_nn}
&& \!\!\!\!\!\!\!\!
\Im m\chi_{nn}({\bm q},\omega_+) = -N_{\rm v} \sum_{ \substack{ {\bm q}',\sigma',\\ \lambda'',\mu''}} \int \frac{d\varepsilon''}{\pi} \big[ n_{\rm F}(\varepsilon'') - n_{\rm F}(\varepsilon''+\omega) \big] 
\nonumber\\
&\times&
\Im m G_{\lambda''}^{(\sigma')}({\bm q}',\varepsilon''_+)
\Im m G_{\mu''}^{(\sigma')}({\bm q'}+{\bm q}, \varepsilon''_++\omega)
\nonumber\\
&\times&
{\cal D}_{\lambda''\mu''}({\bm q}',{\bm q}'+{\bm q}) {\cal D}_{\mu''\lambda''}({\bm q}'+{\bm q},{\bm q}') 
~.
\nonumber\\
\end{eqnarray}
We put Eq.~(\ref{eq:app_Im_chi_nn}) into Eq.~(\ref{eq:app_Im_W}) and then back into Eq.~(\ref{eq:app_taum1}), and we get
\begin{eqnarray} \label{eq:app_taum1_2}
&& \!\!\!\!\!\!\!\!
\frac{1}{\tau_{\rm ee}} = 2N_{\rm v} \sum_{{\bm k}',{\bm q}',\sigma'}\sum_{\substack{ \lambda,\lambda',\\ \lambda'',\mu''}} \int_{-\infty}^{\infty} d\varepsilon \int_{-\infty}^{\infty} \frac{d\varepsilon'}{\pi} \int_{-\infty}^{\infty} \frac{d\varepsilon''}{\pi}
\frac{\partial n_{\rm F}(\varepsilon)}{\partial \varepsilon} 
\nonumber\\
&\times&
\big[ n_{\rm F}(\varepsilon') + n_{\rm B}(\varepsilon'-\varepsilon)\big]
\big[ n_{\rm F}(\varepsilon''+\varepsilon) - n_{\rm F}(\varepsilon''+\varepsilon') \big] 
\nonumber\\
&\times&
|W({\bm k}-{\bm k}',\varepsilon'_+)|
\Im m G^{(\sigma)}({\bm k}',\varepsilon'_+)
\Im m G_{\lambda''}^{(\sigma')}({\bm q}'-{\bm k},\varepsilon''_++\varepsilon)
\nonumber\\
&\times&
\Im m G_{\mu''}^{(\sigma')}({\bm q'}-{\bm k}', \varepsilon''+\varepsilon'_+)
{\cal D}_{\lambda\lambda'}({\bm k},{\bm k}') {\cal D}_{\lambda'\lambda}({\bm k}',{\bm k})
\nonumber\\
&\times&
{\cal D}_{\lambda''\mu''}({\bm q}'-{\bm k},{\bm q}'-{\bm k}')
{\cal D}_{\mu''\lambda''}({\bm q}'-{\bm k}',{\bm q}'-{\bm k}) 
~.
\end{eqnarray}
We now use the fact that
\begin{eqnarray} \label{eq:nB_nF_approx}
{\cal N} &=&
\frac{\partial n_{\rm F}(\varepsilon)}{\partial \varepsilon} \big[n_{\rm F}(\varepsilon') + n_{\rm B}(\varepsilon'-\varepsilon)\big] 
\nonumber\\
&\times&
\big[ n_{\rm F}(\varepsilon''+\varepsilon) - n_{\rm F}(\varepsilon''+\varepsilon') \big] 
\nonumber\\
&=&
\frac{\partial n_{\rm B}(\varepsilon'')}{\partial \varepsilon''} \big[ n_{\rm F}(\varepsilon+\varepsilon'') - n_{\rm F}(\varepsilon) \big] 
\nonumber\\
&\times&
\big[ n_{\rm F}(\varepsilon'+\varepsilon'') - n_{\rm F}(\varepsilon') \big]
\nonumber\\
&\to&
\varepsilon''^2 \frac{\partial n_{\rm B}(\varepsilon'')}{\partial \varepsilon''} \frac{\partial n_{\rm F}(\varepsilon)}{\partial \varepsilon} \frac{\partial n_{\rm F}(\varepsilon')}{\partial \varepsilon'} 
~.
\end{eqnarray}
In evaluating an integral of the form 
\begin{equation}
{\cal I} = \int_{-\infty}^{\infty} d\varepsilon'' \frac{\partial n_{\rm B}(\varepsilon'')}{\partial \varepsilon''} \varepsilon''^2 f(\varepsilon'')
~,
\end{equation}
where $f(\varepsilon'')$ is some smooth function of its argument, we exploit the fact that the weighting function $\varepsilon''^2\partial n_{\rm B}(\varepsilon'')/\partial \varepsilon''$  is strongly peaked at $\varepsilon''=0$ and its width scales with $k_{\rm B}^2 T^2/\varepsilon_{\rm F}$.  This does not mean, however, that one can simply replace $f(\varepsilon'')$ by $f(0)$.  Such a crude approximation would introduce a spurious divergence in the quasiparticle decay rate, because it spoils the subtle cancellation between two infinities which occur (i) in the phase space of the collinear scattering~\cite{Muller_prl_2009,Tomadin_prb_2013} and (ii) in the screening of e-e interactions. Both divergences are connected to the linear-in-momentum energy dispersion of massless Dirac fermions. The cancellation occurs as long as the argument of the function $f(\varepsilon)$ is finite.   To take this into account we approximate
\begin{equation} \label{eq:energy_approx}
\int_{-\infty}^{\infty} d\varepsilon'' \frac{\partial n_{\rm B}(\varepsilon'')}{\partial \varepsilon''} \varepsilon''^2 f(\varepsilon'') = -\frac{2 \pi^2 (k_{\rm B} T)^2}{3} f({\bar \varepsilon}) + {\cal O}(T^4)\,,
\end{equation}
where ${\bar \varepsilon}$ can be estimated as
\begin{equation} \label{eq:SM_epsilon_bar}
{\bar \varepsilon} = \frac{1}{2} \sqrt{-\frac{3}{2\pi^2 (k_{\rm B} T)^2} \int_{-\infty}^{+\infty} d\varepsilon~ \varepsilon^4  \frac{\partial n_{\rm B}(\varepsilon)}{\partial \varepsilon} } = {\bar T} \varepsilon_{\rm F}\,.
\end{equation}
Here we have defined ${\bar T} = \zeta k_{\rm B} T/\varepsilon_{\rm F}$ and $\zeta= \pi/\sqrt{5}$.  The factor $-3/[2\pi^2 (k_{\rm B} T)^2]$ normalizes the weight of the function $\varepsilon^2\partial n_{\rm B}(\varepsilon)/\partial \varepsilon$ to one. We have thus taken ${\bar \varepsilon}$ to be half of the variance of the distribution $\varepsilon^2 \partial n_{\rm B}(\varepsilon)/\partial \varepsilon$.
Eq.~(\ref{eq:energy_approx}) shows the crucial approximation that distinguishes our results for the quasiparticle lifetime from those of Ref.~\onlinecite{Polini_QP_lifetime}. There the authors, although starting from the same GW approximation for the self-energy and deriving an expression equivalent to Eq.~(\ref{eq:app_taum1_2}), did not approximate the final result according to Eq.~(\ref{eq:energy_approx}). The latter allows us to reduce the number of numerical integrations to be performed. With this approximation we finally get the quasiparticle lifetime at the Fermi surface:
\begin{eqnarray} \label{eq:app_taum1_3}
&& \!\!\!\!\!\!\!\!
\frac{1}{\tau_{\rm ee}} = -\frac{4}{3} N_{\rm v} (k_{\rm B} T)^2 \sum_{{\bm k}',{\bm q}',\sigma'}
|W({\bm k}-{\bm k}',{\bar \varepsilon}_+)|
\nonumber\\
&\times&
\Im m G^{(\sigma)}_+({\bm k}',{\bar \varepsilon}_+)
\Im m G_{+}^{(\sigma')}({\bm q}'-{\bm k},0^+)
\nonumber\\
&\times&
\Im m G_{+}^{(\sigma')}({\bm q'}-{\bm k}', {\bar \varepsilon}_+)
{\cal D}_{++}({\bm k},{\bm k}') {\cal D}_{++}({\bm k}',{\bm k})
\nonumber\\
&\times&
{\cal D}_{++}({\bm q}'-{\bm k},{\bm q}'-{\bm k}')
{\cal D}_{++}({\bm q}'-{\bm k}',{\bm q}'-{\bm k}) 
~.
\nonumber\\
\end{eqnarray}
Shifting ${\bm k}' \to {\bm k}-{\bm q}''$ and ${\bm q}' \to {\bm k}-{\bm k}''$ we get
\begin{eqnarray} \label{eq:app_taum1_4}
&& \!\!\!\!\!\!\!\!
\frac{1}{\tau_{\rm ee}} = -\frac{4}{3} N_{\rm v} (k_{\rm B} T)^2 \sum_{{\bm k}'',{\bm q}'',\sigma'}
|W({\bm q}'',{\bar \varepsilon}_+)|
\nonumber\\
&\times&
\Im m G^{(\sigma)}_+({\bm k}-{\bm q}'',{\bar \varepsilon}_+)
\Im m G_{+}^{(\sigma')}({\bm k}''-{\bm q}'', {\bar \varepsilon}_+)
\nonumber\\
&\times&
\Im m G_{+}^{(\sigma')}({\bm k}'',0^+)
{\cal D}_{++}({\bm k},{\bm k}-{\bm q}'') {\cal D}_{++}({\bm k}-{\bm q}'',{\bm k})
\nonumber\\
&\times&
{\cal D}_{++}({\bm k}'',{\bm k}''-{\bm q}'')
{\cal D}_{++}({\bm k}''-{\bm q}'',{\bm k}'') 
~,
\nonumber\\
\end{eqnarray}
which can be recasted into the following Fermi-golden-rule form
\begin{eqnarray} \label{eq:inverse_lifetime_2}
\frac{1}{\tau_{\rm qp}^{\rm ee}} 
&=&
\frac{4\pi}{3} (k_{\rm B} T)^2 \sum_{{\bm q}}
|W({\bm q},{\bar \varepsilon})|^2
\frac{\Im m \chi_{nn}({\bm q}, {\bar \varepsilon})}{{\bar \varepsilon}}
\nonumber\\
&\times&
\Im m \big[G^{({\rm R},\sigma)}_{+}({\bm k} - {\bm q},-{\bar \varepsilon})\big]
\frac{1+\cos(\varphi_{{\bm k}-{\bm q}}-\varphi_{{\bm k}})}{2}
~.
\nonumber\\
\end{eqnarray}
Note that in these equation we do not sum over the spin index $\sigma$. Since the system is spin-unpolarized the lifetimes at the Fermi surface of the two spin populations coincide.

\section{Analytical continuation of the response function}
\label{app:analytical_continuation}
To analytically continue $i\omega_m \to \omega+i\eta$ in Eq.~(\ref{eq:chi_jj_def}) we define the function $f(i\varepsilon,i\varepsilon+i\omega)$ such that
\begin{eqnarray} \label{eq:chi_jj_f}
\chi_{j^{(\sigma)}_{\alpha} j^{(\sigma')}_{\beta}} ({\bm q}, i\omega_m) &\equiv& k_{\rm B} T \sum_{\varepsilon_n} f(i\varepsilon_n, i\varepsilon_n+i\omega_m)
\nonumber\\
&=& -\oint \frac{dz}{2\pi i} n_{\rm F}(z) f(z,z+i\omega_m)
~,
\nonumber\\
\end{eqnarray}
where we suppress for brevity all its momentum dependence. The contour of integration in Eq.~(\ref{eq:chi_jj_f}) is chosen in such a way as to encircle all the poles of $n_{\rm F}(z)$ and to exclude all the branch cuts of $f(z,z+i\omega_m)$, which occur for $\Im m (z) = 0$ and $\Im m (z+i\omega_m) = 0$. We transform the contour-integration in an integration over the branch cuts and we then perform the analytical continuation $i\omega_m \to \omega + i\eta$. We get
\begin{eqnarray} \label{eq:chi_jj_final}
&& \!\!\!\!\!\!\!\!
\chi_{j^{(\sigma)}_{\alpha} j^{(\sigma')}_{\beta}} ({\bm q}, \omega) = -\int \frac{d\varepsilon}{2\pi i} \Big\{\big[n_{\rm F} (\varepsilon+\omega) - n_{\rm F} (\varepsilon)\big] 
\nonumber\\
&\times&
\big[f(\varepsilon_-, \varepsilon_++\omega) - f(\varepsilon_-, \varepsilon_-+\omega)\big]
\nonumber\\
&+&
n_{\rm F}(\varepsilon) \big[f(\varepsilon_+, \varepsilon_++\omega) - f(\varepsilon_-, \varepsilon_-+\omega)\big] \Big\}
~.
\end{eqnarray}
Here $\varepsilon_\pm = \varepsilon\pm i\eta$. After the analytical continuation, $G_{\lambda}^{(\sigma)}({\bm k},\varepsilon_+) = G^{({\rm R},\sigma)}_{\lambda}({\bm k},\varepsilon)$ and $G_{\lambda}^{(\sigma)}({\bm k},\varepsilon_-) = G^{({\rm A},\sigma)}_{\lambda}({\bm k},\varepsilon)$. Here $G^{({\rm A},\sigma)}_{\lambda}({\bm k}, \varepsilon)$ [$G^{({\rm R},\sigma)}_{\lambda}({\bm k}, \varepsilon)$] represents the advanced [retarded] Green's function. 

Note that the square brackets in the last line of Eq.~(\ref{eq:chi_jj_final}) contain a purely imaginary quantity, which (being divided by the imaginary unit) gives a purely real contribution to $\chi_{j^{(\sigma)}_{\alpha} j^{(\sigma')}_{\beta}} ({\bm q}, \omega)$. Note also that $f(\varepsilon_-, \varepsilon_++\omega)$ contains the product of a retarded and an advanced Green's function, whereas in $f(\varepsilon_-, \varepsilon_-+\omega)$ and $f(\varepsilon_+, \varepsilon_++\omega)$ both Green's functions are either advanced or retarded. The last two functions [$f(\varepsilon_-, \varepsilon_-+\omega)$ and $f(\varepsilon_+, \varepsilon_++\omega)$] have all the poles on the same side of the complex plane. Note however that, as usual~\cite{Bruus_and_Flensberg}, we can exploit this property only performing the integral over the band energies. We thus get that $f(\varepsilon_-, \varepsilon_++\omega)$ gives the dominant contribution when $\varepsilon_{\rm F} \tau_{\rm ee} \gg 1$. Since we are interested in the Fermi liquid regime, in what follows we will retain only this term. Eq.~(\ref{eq:chi_jj_final}) thus becomes
\begin{eqnarray} \label{eq:chi_jj_final_omega_finite}
&& \!\!\!\!\!\!\!\!
\chi_{j^{(\sigma)}_{\alpha} j^{(\sigma')}_{\beta}} ({\bm q}, \omega) = - N_{\rm v} \sum_{{\bm k}, \lambda, \lambda'} \int \frac{d\varepsilon}{2\pi i} \big[n_{\rm F} (\varepsilon+\omega) - n_{\rm F} (\varepsilon)\big]
\nonumber\\
&\times&
G^{({\rm A},\sigma)}_{\lambda}({\bm k}_-, \varepsilon) \Lambda^{(0,\sigma)}_{\lambda\lambda',\alpha}({\bm k}_-,{\bm k}_+)
G^{({\rm R},\sigma)}_{\lambda'}({\bm k}_+, \varepsilon + \omega)
\nonumber\\
&\times&
\Lambda^{(\sigma\sigma')}_{\lambda'\lambda,\beta} ({\bm k}_+, \varepsilon_+ + \omega,{\bm k}_-,\varepsilon_-)
~,
\end{eqnarray}
which coincides with Eq.~(\ref{eq:R_chi_jj_final_omega_finite}).

\section{Analytical continuation of the Bethe-Salpeter equation} \label{sect:fpm}
In this section we guide the reader through the long and complicated calculation of the vertex correction to the thermal conductivity. We analytically continue the three contributions $\Lambda^{(i,\sigma\sigma')}_{\lambda'\lambda,\beta} ({\bm k}_+, i\varepsilon_n + i\omega_m,{\bm k}_-,i\varepsilon_n)$ ($i=1,\ldots,3$), defined in Eqs.~(\ref{eq:Lambda_1_2})-(\ref{eq:W_3}) to real frequencies. 
In the Fermi-liquid regime we consider only the dominant contribution to the dressed vertex, to be used in combination with the product of the retarded and advanced Green's functions that appears in Eq.~(\ref{eq:chi_jj_final_omega_finite}). From this we see that the analytic continuation of $\Lambda^{(i,\sigma\sigma')}_{\lambda'\lambda,\beta} ({\bm k}_+, i\varepsilon_n + i\omega_m,{\bm k}_-,i\varepsilon_n)$ is done with the prescriptions $i\omega_m\to \omega_+$, $i\varepsilon_n\to \varepsilon_-$, $i\varepsilon_n+i\omega_m \to \varepsilon_++\omega$.

\subsection{Analytical continuation of Eq.~(\ref{eq:Lambda_1_2})}
We define the function $g(i\varepsilon_{n'},i\varepsilon_{n'} + i\omega_m,i\varepsilon_{n'}-i\varepsilon_n)$ such that
\begin{eqnarray} \label{eq:Lambda_1_2_f}
&& \!\!\!\!\!\!\!\!
\Lambda^{(1,2)} ({\bm k}_+, i\varepsilon_n + i\omega_m,{\bm k}_-,i\varepsilon_n) 
\nonumber\\
&\equiv&
-k_{\rm B} T \sum_{\varepsilon_{n'}} g(i\varepsilon_{n'},i\varepsilon_{n'} + i\omega_m,i\varepsilon_{n'}-i\varepsilon_n)
\nonumber\\
&=&
\oint \frac{dz}{2\pi i} n_{\rm F}(z) g(z,z + i\omega_m,z-i\varepsilon_n)
~.
\end{eqnarray}
Here and in what follows we suppress for brevity all the spin, band and spatial indices of the dressed vertex. As usual, we transform the sum over the poles of $n_{\rm F}(z)$ in an integration over the branch cuts of $g(z,z + i\omega_m,z-i\varepsilon_n)$. We then perform the analytic continuations with the prescription $i\omega_m\to \omega_+$, $i\varepsilon_n\to \varepsilon_-$, $i\varepsilon_n+i\omega_m \to \varepsilon_++\omega$. After some lengthy but straightforward algebra we get
\begin{eqnarray} \label{eq:Lambda_1_2_f_BC}
&& \!\!\!\!\!\!\!\!
\Lambda^{(1,2)} ({\bm k}_+, \varepsilon_+ + \omega,{\bm k}_-,\varepsilon_-) =
\int \frac{d\varepsilon'}{2\pi i} \Big\{ 
n_{\rm F}(\varepsilon') 
\nonumber\\
&\times&
\big[ g(\varepsilon'_+,\varepsilon'_+ + \omega,\varepsilon'_+-\varepsilon) - g(\varepsilon'_-,\varepsilon'_+ + \omega,\varepsilon'_+-\varepsilon)\big]
\nonumber\\
&+&
n_{\rm F}(\varepsilon') \big[ g(\varepsilon'_--\omega,\varepsilon'_+ ,\varepsilon'_--\varepsilon-\omega) 
\nonumber\\
&-& g(\varepsilon'_--\omega,\varepsilon'_-,\varepsilon'_--\varepsilon-\omega)\big]
\nonumber\\
&-&
n_{\rm B}(\varepsilon')
\big[ g(\varepsilon'_-+\varepsilon,\varepsilon'_+ +\varepsilon + \omega,\varepsilon'_+) 
\nonumber\\
&-&
g(\varepsilon'_-+\varepsilon,\varepsilon'_+ + \varepsilon+\omega,\varepsilon'_-)\big]
\Big\}
~.
\end{eqnarray}
We now shift $\varepsilon'\to \varepsilon'+\omega$ in the third and fourth lines of Eq.~(\ref{eq:Lambda_1_2_f_BC}). We note that we can safely take the limit $\omega \to 0$ in $n_{\rm F}(\varepsilon'+\omega)$. Note also that $g(\varepsilon'_+,\varepsilon'_+ + \omega,\varepsilon'_+-\varepsilon)$ and $g(\varepsilon'_--\omega,\varepsilon'_-,\varepsilon'_--\varepsilon-\omega)$ have the poles on the same side of the complex plane, and therefore can be neglected~\cite{Bruus_and_Flensberg} in the limit of $\varepsilon_{\rm F} \tau_{\rm ee} \gg 1$. Shifting $\varepsilon'\to \varepsilon'+\varepsilon$ in the last two lines of Eq.~(\ref{eq:Lambda_1_2_f_BC}) we readily obtain
\begin{eqnarray} \label{eq:Lambda_1_2_f_final}
&&
\Lambda^{(1,2)} ({\bm k}_+, \varepsilon_++\omega,{\bm k}_-,\varepsilon_-) =
\sum_{{\bm k}',\sigma''} \sum_{\mu,\mu'} \int \frac{d\varepsilon'}{2\pi i} 
\nonumber\\
&& \times
\Big[ W^{(1,2,\sigma\sigma'')}_{\lambda\lambda'\mu\mu'}({\bm k}',{\bm k},\varepsilon'_--\varepsilon) - W^{(1,2,\sigma\sigma'')}_{\lambda\lambda'\mu\mu'}({\bm k}',{\bm k},\varepsilon'_+-\varepsilon) \Big]
\nonumber\\
&& \times
\big[n_{\rm F}(\varepsilon') + n_{\rm B}(\varepsilon'-\varepsilon)\big]
G^{({\rm R},\sigma'')}_{\mu'}({\bm k}'_+,\varepsilon'+\omega) 
\nonumber\\
&& \times
G^{({\rm A},\sigma'')}_{\mu}({\bm k}'_-,\varepsilon')
\Lambda^{(\sigma''\sigma')}_{\mu'\mu,\beta} ({\bm k}'_+, \varepsilon'_++\omega, {\bm k}'_-,\varepsilon'_-)
~.
\end{eqnarray}
It remains to determine $W^{(1,2,\sigma\sigma'')}_{\lambda\lambda'\mu\mu'}({\bm k}',{\bm k},\varepsilon'_\pm-\varepsilon)$. Eq.~(\ref{eq:Lambda_1_2_f_final}) implies that we have to analytically continue the functions $W^{(1,2)}$ with the prescription $i\varepsilon_{n'} \to \varepsilon'_-$ and $i\varepsilon_{n'} + i\omega_m \to \varepsilon'_+ + \omega$.

\subsubsection{The analytical continuation of Eq.~(\ref{eq:W_1})}
We now perform the analytical continuation of Eq.~(\ref{eq:W_1}) with the prescription $i\omega_m\to \omega_+$, $i\varepsilon_n\to \varepsilon_-$, $i\varepsilon_n+i\omega_m \to \varepsilon_++\omega$, $i\varepsilon_{n'} \to \varepsilon'_-$, and $i\varepsilon_{n'} + i\omega_m \to \varepsilon'_+ + \omega$. As shown in Eq.~(\ref{eq:Lambda_1_2_f_final}), we need to calculate the function $W^{(1,\sigma\sigma'')}_{\lambda\lambda'\mu\mu'}({\bm k},{\bm k'},\varepsilon_-'-\varepsilon) - W^{(1,\sigma\sigma'')}_{\lambda\lambda'\mu\mu'}({\bm k},{\bm k'},\varepsilon_+'-\varepsilon)$, which reads
\begin{eqnarray} \label{eq:W_1_A-R}
&& \!\!\!\!\!\!\!\!
W^{(1,\sigma\sigma'')}_{\lambda\lambda'\mu\mu'}({\bm k},{\bm k'},\varepsilon_-'-\varepsilon) - W^{(1,\sigma\sigma'')}_{\lambda\lambda'\mu\mu'}({\bm k},{\bm k'},\varepsilon_+'-\varepsilon) = \delta_{\sigma\sigma''} 
\nonumber\\
&\times&
\big[W({\bm k}-{\bm k}',\varepsilon_-'-\varepsilon) - W({\bm k}-{\bm k}',\varepsilon_+'-\varepsilon)\big]
\nonumber\\
&\times&
{\cal D}_{\lambda'\mu'}({\bm k}_+,{\bm k}'_+) {\cal D}_{\mu\lambda}({\bm k}'_-,{\bm k}_-)
\nonumber\\
&=&
- 2 i\delta_{\sigma\sigma''} \Im m \big[W({\bm k}-{\bm k}',\varepsilon_+'-\varepsilon)\big]
\nonumber\\
&\times&
{\cal D}_{\lambda'\mu'}({\bm k}_+,{\bm k}'_+) {\cal D}_{\mu\lambda}({\bm k}'_-,{\bm k}_-)
\nonumber\\
&=& \!\!\!
-2 i \delta_{\sigma\sigma''} |W({\bm k}-{\bm k}',\varepsilon'-\varepsilon)|^2 \Im m \big[\chi_{nn}^{(0)}({\bm k}-{\bm k}',\varepsilon_+'-\varepsilon)\big]
\nonumber\\
&\times&
{\cal D}_{\lambda'\mu'}({\bm k}_+,{\bm k}'_+) {\cal D}_{\mu\lambda}({\bm k}'_-,{\bm k}_-)
~.
\end{eqnarray}
The imaginary part of the density-density response function was given in Eq.~(\ref{eq:app_Im_chi_nn}). After some straightforward manipulation, we get 
\begin{eqnarray} \label{eq:chi_nn_final}
&&
\Im m \big[\chi_{nn} ({\bm k}-{\bm k}', \varepsilon'_+-\varepsilon)\big] = 
2 N_{\rm v} \sum_{{\bm q}',\sigma'''} \sum_{\lambda'',\mu''} \int \frac{d\omega'}{2\pi} 
\nonumber\\
&& \times
\big[n_{\rm F} (\omega'+\varepsilon') - n_{\rm F} (\omega'+\varepsilon)\big] 
\nonumber\\
&& \times
\Im m \Big[G^{({\rm R},\sigma''')}_{\lambda''}({\bm q}'-{\bm k},\omega'+\varepsilon)\Big]
\nonumber\\
&& \times
\Im m\Big[G^{({\rm R},\sigma''')}_{\mu''}({\bm q'}-{\bm k}', \omega'+\varepsilon')\Big]
\nonumber\\
&& \times
{\cal D}_{\lambda''\mu''}({\bm q}'-{\bm k},{\bm q}'-{\bm k}') {\cal D}_{\mu''\lambda''}({\bm q}'-{\bm k}',{\bm q}'-{\bm k}) 
~.
\end{eqnarray}
Putting Eq.~(\ref{eq:chi_nn_final}) into Eq.~(\ref{eq:W_1_A-R}) we finally find
\begin{eqnarray} \label{eq:W_1_A-R_2}
&&
W^{(1,\sigma\sigma'')}_{\lambda\lambda'\mu\mu'}({\bm k},{\bm k'},\varepsilon_-'-\varepsilon) - W^{(1,\sigma\sigma'')}_{\lambda\lambda'\mu\mu'}({\bm k},{\bm k'},\varepsilon_+'-\varepsilon)
\nonumber\\
&& =
4 N_{\rm v} 
\delta_{\sigma\sigma''} |W({\bm k}-{\bm k}',\varepsilon'-\varepsilon)|^2
\! \sum_{{\bm q}',\sigma'''} \! \sum_{\lambda'',\mu''}
\nonumber\\
&& \times
\int \frac{d\omega'}{2\pi i} \big[n_{\rm F} (\omega'+\varepsilon') - n_{\rm F} (\omega'+\varepsilon)\big] 
\nonumber\\
&& \times
\Im m \Big[G^{({\rm R},\sigma''')}_{\lambda''}({\bm q}'-{\bm k},\omega'+\varepsilon)\Big]
\nonumber\\
&& \times
\Im m\Big[G^{({\rm R},\sigma''')}_{\mu''}({\bm q'}-{\bm k}', \omega'+\varepsilon')\Big]
\nonumber\\
&& \times
{\cal D}_{\lambda'\mu'}({\bm k}_+,{\bm k}'_+) {\cal D}_{\mu\lambda}({\bm k}'_-,{\bm k}_-)
\nonumber\\
&& \times
{\cal D}_{\lambda''\mu''}({\bm q}'-{\bm k},{\bm q}'-{\bm k}') {\cal D}_{\mu''\lambda''}({\bm q}'-{\bm k}',{\bm q}'-{\bm k}) 
~.
\nonumber\\
\end{eqnarray}

\subsubsection{The analytical continuation of Eq.~(\ref{eq:W_2})}
We now turn to the analytical continuation of Eq.~(\ref{eq:W_2}) with the prescription $i\omega_m\to \omega_+$, $i\varepsilon_n\to \varepsilon_-$, $i\varepsilon_n+i\omega_m \to \varepsilon_++\omega$, $i\varepsilon_{n'} \to \varepsilon'_-$, and $i\varepsilon_{n'} + i\omega_m \to \varepsilon'_+ + \omega$. This time we define 
\begin{eqnarray} \label{eq:W_1_2_f}
&& \!\!\!\!\!\!\!\!
W^{(2,\sigma\sigma'')}_{\lambda\lambda'\mu\mu'} ({\bm k}',{\bm k},i\varepsilon_{n'}-i\varepsilon_n) \equiv 
\oint \frac{dz}{2\pi i} n_{\rm B}(z) 
\nonumber\\
&\times&
w_2 (z, z-i\omega_m,i\varepsilon_{n}+i\omega_m -z, i\varepsilon_{n'}+i\omega_m -z)
~.
\nonumber\\
\end{eqnarray}
Integrating over the branch cuts of $w_2 (z, z-i\omega_m,i\varepsilon_{n}+i\omega_m -z, i\varepsilon_{n'}+i\omega_m -z)$, and performing the analytical continuations as stated before, we get
\begin{eqnarray} \label{eq:W_1_2_f_BC}
&& \!\!\!\!\!\!\!\!
W^{(2,\sigma\sigma'')}_{\lambda\lambda'\mu\mu'} ({\bm k}',{\bm k},\varepsilon'_\pm-\varepsilon) =
\int \frac{d\omega'}{2\pi i} \Big\{
n_{\rm B}(\omega') 
\nonumber\\
&\times&
\big[w_2 (\omega'_+, \omega'_--\omega,\varepsilon + \omega -\omega'_-, \varepsilon'+\omega -\omega'_-) 
\nonumber\\
&-&
w_2 (\omega'_-, \omega'_--\omega,\varepsilon + \omega -\omega'_-, \varepsilon'+\omega -\omega'_-)\big]
\nonumber\\
&+&
n_{\rm B}(\omega') \big[w_2 (\omega'_++\omega, \omega'_+,\varepsilon -\omega'_+, \varepsilon' -\omega'_+) 
\nonumber\\
&-&
w_2 (\omega'_++\omega, \omega'_-,\varepsilon -\omega'_+, \varepsilon' -\omega'_+) \big]
\nonumber\\
&-&
n_{\rm F}(\omega') \big[w_2 (\omega'_++\varepsilon+\omega, \omega'_-+\varepsilon, -\omega'_+, \varepsilon'-\varepsilon -\omega'_\mp) 
\nonumber\\
&-&
w_2 (\omega'_++\varepsilon+\omega, \omega'_-+\varepsilon, -\omega'_-, \varepsilon'-\varepsilon -\omega'_\mp) \big]
\nonumber\\
&-&
n_{\rm F}(\omega') \big[w_2 (\omega'_++\varepsilon'+\omega, \omega'_-+\varepsilon', \varepsilon-\varepsilon' -\omega'_\pm, -\omega'_+) 
\nonumber\\
&-&
w_2 (\omega'_++\varepsilon'+\omega, \omega'_-+\varepsilon', \varepsilon-\varepsilon' -\omega'_\pm, -\omega'_-) \big]
\Big\}
~.
\nonumber\\
\end{eqnarray}
Note that the terms on the r.h.s. of Eq.~(\ref{eq:W_1_2_f_BC}) proportional to $n_{\rm B}(\omega')$ are identical in both $W^{(2,\sigma\sigma'')}_{\lambda\lambda'\mu\mu'} ({\bm k}',{\bm k},\varepsilon'_\pm-\varepsilon)$ and thus vanish when the difference is taken. We thus neglect them in what follows. Eq.~(\ref{eq:W_1_2_f_BC}) reduces to
\begin{eqnarray} \label{eq:W_1_2_f_final}
&& \!\!\!\!\!\!\!\!
W^{(2,\sigma\sigma'')}_{\lambda\lambda'\mu\mu'} ({\bm k}',{\bm k},\varepsilon'_\pm-\varepsilon) =
-\int \frac{d\omega'}{2\pi i} \Big\{
n_{\rm F}(\omega'-\varepsilon-\omega)
\nonumber\\
&\times&
\big[w_2 (\omega'_+, \omega'_--\omega, \varepsilon+\omega-\omega'_+, \varepsilon' +\omega-\omega'_\mp) 
\nonumber\\
&-&
w_2 (\omega'_+, \omega'_--\omega, \varepsilon+\omega-\omega'_-, \varepsilon'+\omega-\omega'_\mp) \big]
\nonumber\\
&+&
n_{\rm F}(\omega'-\varepsilon'-\omega)
\nonumber\\
&\times&
\big[w_2 (\omega'_++, \omega'_--\omega, \varepsilon +\omega-\omega'_\pm, \varepsilon'+\omega-\omega'_+) 
\nonumber\\
&-&
w_2 (\omega'_+, \omega'_--\omega, \varepsilon +\omega-\omega'_\pm, \varepsilon'+\omega-\omega'_-) \big]
\Big\}
~.
\nonumber\\
\end{eqnarray}
Finally,
\begin{eqnarray} \label{eq:W_1_2_A-R}
&& \!\!\!\!\!\!\!\!
W^{(2,\sigma\sigma'')}_{\lambda\lambda'\mu\mu'} (\varepsilon'_--\varepsilon) - W^{(2,\sigma\sigma'')}_{\lambda\lambda'\mu\mu'} (\varepsilon'_+-\varepsilon) = 
\int \frac{d\omega'}{2\pi i} 
\nonumber\\
&\times&
\big[n_{\rm F}(\omega'-\varepsilon-\omega) - n_{\rm F}(\omega'-\varepsilon'-\omega)\big]
\nonumber\\
&\times&
\big[
w_2 (\omega'_+, \omega'_--\omega, \varepsilon+\omega-\omega'_+, \varepsilon'+\omega -\omega'_-) 
\nonumber\\
&-&
w_2 (\omega'_+, \omega'_--\omega, \varepsilon+\omega-\omega'_-, \varepsilon'+\omega-\omega'_-) 
\nonumber\\
&-&
w_2 (\omega'_+, \omega'_--\omega, \varepsilon+\omega-\omega'_+, \varepsilon'+\omega -\omega'_+) 
\nonumber\\
&+&
w_2 (\omega'_+, \omega'_--\omega, \varepsilon+\omega-\omega'_-, \varepsilon'+\omega-\omega'_+)\big]
\nonumber\\
&=&
4 N_{\rm v} \sum_{{\bm q}',\lambda'',\mu''} \int \frac{d\omega'}{2\pi i} 
W({\bm q}',\omega'_+) W({\bm q}',\omega'_--\omega)
\nonumber\\
&\times&
\big[n_{\rm F}(\omega'-\varepsilon-\omega) - n_{\rm F}(\omega'-\varepsilon'-\omega)\big]
\nonumber\\
&\times&
\Im m\big[ G^{({\rm R},\sigma)}_{\lambda''}({\bm k}_+-{\bm q}',\varepsilon+\omega-\omega') \big]
\nonumber\\
&\times&
\Im m\big[ G^{({\rm R},\sigma'')}_{\mu''}({\bm k}'_+-{\bm q}',\varepsilon'+\omega -\omega') \big]
\nonumber\\
&\times&
{\cal D}_{\lambda'\lambda''}({\bm k}_+,{\bm k}_+-{\bm q}')
{\cal D}_{\lambda''\lambda}({\bm k}_+-{\bm q}',{\bm k}_-)
\nonumber\\
&\times&
{\cal D}_{\mu\mu''}({\bm k}'_-,{\bm k}'_+-{\bm q}')
{\cal D}_{\mu''\mu'}({\bm k}'_+-{\bm q}',{\bm k}'_+)
~.
\end{eqnarray}
We can now take the limit $v_{\rm F} q\ll \omega \ll \varepsilon_{\rm F}$, and we get
\begin{eqnarray} \label{eq:W_2_A-R_2}
&& W^{(2,\sigma\sigma'')}_{\lambda\lambda'\mu\mu'} (\varepsilon'_--\varepsilon) - W^{(2,\sigma\sigma'')}_{\lambda\lambda'\mu\mu'} (\varepsilon'_+-\varepsilon) 
= 4 N_{\rm v}
\sum_{{\bm q}',\lambda'',\mu''}
\nonumber\\
&& \times
\int \frac{d\omega'}{2\pi i} 
\big|W({\bm q}',\omega')\big|^2 \big[n_{\rm F}(\omega'-\varepsilon) - n_{\rm F}(\omega'-\varepsilon')\big]
\nonumber\\
&& \times
\Im m \big[ G^{({\rm R},\sigma)}_{\lambda''}({\bm k}-{\bm q}',\varepsilon-\omega') \big]
\nonumber\\
&& \times
\Im m \big[ G^{({\rm R},\sigma'')}_{\mu''}({\bm k}'-{\bm q}',\varepsilon' -\omega') \big]
\nonumber\\
&& \times
{\cal D}_{\lambda'\lambda''}({\bm k},{\bm k}-{\bm q}')
{\cal D}_{\lambda''\lambda}({\bm k}-{\bm q}',{\bm k})
\nonumber\\
&& \times
{\cal D}_{\mu\mu''}({\bm k}',{\bm k}'-{\bm q}')
{\cal D}_{\mu''\mu'}({\bm k}'-{\bm q}',{\bm k}')
~.
\nonumber\\
\end{eqnarray}

\subsection{Analytical continuation of Eq.~(\ref{eq:Lambda_3})}
We define $h(i\varepsilon_{n'}, i\varepsilon_{n'} + i\omega_m, i\varepsilon_{n'}+i\varepsilon_{n} +i\omega_m)$ such that
\begin{eqnarray} \label{eq:Lambda_3_f}
&& \!\!\!\!\!\!\!\!
\Lambda^{(3)} ({\bm k}_+, i\varepsilon_n + i\omega_m,{\bm k}_-,i\varepsilon_n)
\nonumber\\
&\equiv&
- k_{\rm B} T \sum_{\varepsilon_{n'}} h(i\varepsilon_{n'}, i\varepsilon_{n'} + i\omega_m, i\varepsilon_{n'}+i\varepsilon_{n} +i\omega_m)
\nonumber\\
&=&
\oint \frac{dz}{2\pi i} n_{\rm F}(z) h(z, z + i\omega_m, z+i\varepsilon_{n} +i\omega_m)
~.
\nonumber\\
\end{eqnarray}
Here and in what follows we suppress for brevity all the spin, band and spatial indices. To perform the analytical continuation we first transform the sum over the poles of $n_{\rm F}(z)$ in an integration over the branch cuts of $h(z, z + i\omega_m, z+i\varepsilon_{n} +i\omega_m)$. We then analytically continue the result, according to the prescription $i\omega_m\to \omega_+$, $i\varepsilon_n\to \varepsilon_-$, $i\varepsilon_n+i\omega_m \to \varepsilon_++\omega$. After some lengthy algebra we get
\begin{eqnarray} \label{eq:Lambda_3_f_BC}
&& \!\!\!\!\!\!\!\!
\Lambda^{(3)} ({\bm k}_+, \varepsilon_+ + \omega,{\bm k}_-,\varepsilon_-) =
\int \frac{d\varepsilon'}{2\pi i} \Big\{
n_{\rm F}(\varepsilon')
\nonumber\\
&\times& 
\big[ h(\varepsilon'_+, \varepsilon'_+ + \omega, \varepsilon'_++\varepsilon +\omega) 
\nonumber\\
&-&
h(\varepsilon'_-, \varepsilon'_+ + \omega, \varepsilon'_++\varepsilon +\omega) \big]
\nonumber\\
&+&
n_{\rm F}(\varepsilon') \big[ h(\varepsilon'_--\omega, \varepsilon'_+, \varepsilon'_-+\varepsilon) 
\nonumber\\
&-&
h(\varepsilon'_--\omega, \varepsilon'_-, \varepsilon'_-+\varepsilon) \big]
\nonumber\\
&-&
n_{\rm B}(\varepsilon') \big[ h(\varepsilon'_- -\varepsilon -\omega, \varepsilon'_+ - \varepsilon, \varepsilon'_+) 
\nonumber\\
&-&
h(\varepsilon'_--\varepsilon -\omega, \varepsilon'_+ -\varepsilon, \varepsilon'_-) \big] \Big\}
~.
\end{eqnarray}
We now shift $\varepsilon'\to \varepsilon'+\omega$ in the third and fourth lines of Eq.~(\ref{eq:Lambda_3_f_BC}), and  we take the limit $\omega\to 0$ in $n_{\rm F}(\varepsilon'+\omega)$. We note that $h(\varepsilon'_+, \varepsilon'_+ + \omega, \varepsilon'_++\varepsilon +\omega)$ and $h(\varepsilon'_--\omega, \varepsilon'_-, \varepsilon'_-+\varepsilon)$ have the poles on the same side of the complex plane, and can be neglected in the limit $\varepsilon_{\rm F} \tau_{\rm ee} \gg 1$. We then shift $\varepsilon'\to \varepsilon'+\varepsilon+\omega$ in the last two lines of Eq.~(\ref{eq:Lambda_3_f_BC}), and we take the limit $\omega \to 0$ in $n_{\rm B}(\varepsilon'+\varepsilon+\omega)$. After these manipulations Eq.~(\ref{eq:Lambda_3_f_BC}) becomes
\begin{eqnarray} \label{eq:Lambda_3_f_final}
&&
\Lambda^{(3)} ({\bm k}_+, \varepsilon_++\omega,{\bm k}_-,\varepsilon_-) =
\sum_{{\bm k}',\sigma''} \sum_{\mu,\mu'} \int \frac{d\varepsilon'}{2\pi i} 
\nonumber\\
&& \times
\Big[W^{(3,\sigma\sigma'')}_{\lambda\lambda'\mu\mu'}({\bm k}, {\bm k}', \varepsilon'_-+\varepsilon) - W^{(3,\sigma\sigma'')}_{\lambda\lambda'\mu\mu'}({\bm k}, {\bm k}', \varepsilon'_++\varepsilon) \Big]
\nonumber\\
&& \times
\big[n_{\rm F}(\varepsilon') + n_{\rm B}(\varepsilon' + \varepsilon)\big]
G^{({\rm R},\sigma'')}_{\mu'}({\bm k}'_+,\varepsilon'+\omega) 
\nonumber\\
&& \times
G^{({\rm A},\sigma'')}_{\mu}({\bm k}'_-,\varepsilon')
\Lambda^{(\sigma''\sigma')}_{\mu'\mu,\beta} ({\bm k}'_+, \varepsilon_++\omega,{\bm k}'_-,\varepsilon'_-)
~.
\end{eqnarray}
It only remains to determine $W^{(3,\sigma\sigma'')}_{\lambda\lambda'\mu\mu'} ({\bm k}'-{\bm k},\varepsilon'_\pm-\varepsilon)$, defined in Eq.~(\ref{eq:W_3}). Eq.~(\ref{eq:Lambda_3_f_final}) implies that we have to analytically continue the functions $W^{(3)}$ for $i\varepsilon_{n'} \to \varepsilon'_-$ and $i\varepsilon_{n'} + i\omega_m \to \varepsilon'_+ + \omega$.

\subsubsection{The analytical continuation of Eq.~(\ref{eq:W_3})}
We now turn to the analytical continuation of Eq.~(\ref{eq:W_3}) with the prescription $i\omega_m\to \omega_+$, $i\varepsilon_n\to \varepsilon_-$, $i\varepsilon_n+i\omega_m \to \varepsilon_++\omega$, $i\varepsilon_{n'} \to \varepsilon'_-$, and $i\varepsilon_{n'} + i\omega_m \to \varepsilon'_+ + \omega$. We define
\begin{eqnarray} \label{eq:W_3_f}
&& \!\!\!\!\!\!\!\!
W^{(3,\sigma\sigma'')}_{\lambda\lambda'\mu\mu'} ({\bm k}',{\bm k},i\varepsilon_{n'}+i\varepsilon_n+i\omega_m) \equiv
\oint \frac{dz}{2\pi i} n_{\rm B}(z)
\nonumber\\
&\times&
w_3(z,z-i\omega_m,i\varepsilon_n +z,i\varepsilon_{n'}+i\omega_m -z)
~.
\end{eqnarray}
Integrating over the branch cuts of $w_3(z,z-i\omega_m,i\varepsilon_n +z,i\varepsilon_{n'}+i\omega_m -z)$, and performing the analytical continuations according to the prescriptions stated before, we get
\begin{eqnarray} \label{eq:W_3_f_BC}
&& \!\!\!\!\!\!\!\!
W^{(3,\sigma\sigma'')}_{\lambda\lambda'\mu\mu'} ({\bm k}',{\bm k},\varepsilon'_\pm+\varepsilon+\omega) =
\int \frac{d\omega'}{2\pi i} \Big\{
n_{\rm B}(\omega') 
\nonumber\\
&\times&
\big[w_3(\omega'_+,\omega'_--\omega,\omega'_- + \varepsilon, \varepsilon'+ \omega -\omega'_-)
\nonumber\\
&-&
w_3(\omega'_-,\omega'_--\omega,\omega'_- + \varepsilon, \varepsilon'+ \omega -\omega'_-)
\big]
\nonumber\\
&+&
n_{\rm B}(\omega') \big[w_3(\omega'_++\omega,\omega'_+,\omega'_++ \varepsilon+\omega, \varepsilon' -\omega'_+)
\nonumber\\
&-&
w_3(\omega'_++\omega,\omega'_-,\omega'_++ \varepsilon+\omega, \varepsilon' -\omega'_+)
\big]
\nonumber\\
&-&
n_{\rm F}(\omega')
\nonumber\\
&\times&
\big[w_3(\omega'_+-\varepsilon,\omega'_--\varepsilon-\omega,\omega'_+, \varepsilon'+\varepsilon+\omega -\omega'_\mp)
\nonumber\\
&-&
w_3(\omega'_+-\varepsilon,\omega'_--\varepsilon-\omega,\omega'_-, \varepsilon'+\varepsilon+\omega -\omega'_\mp)
\big]
\nonumber\\
&-&
n_{\rm F}(\omega')
\nonumber\\
&\times&
\big[w_3(\omega'_++\varepsilon'+\omega,\omega'_-+\varepsilon',\omega'_\pm + \varepsilon+\varepsilon'+\omega,  -\omega'_+)
\nonumber\\
&-&
w_3(\omega'_++\varepsilon'+\omega,\omega'_-+\varepsilon',\omega'_\pm + \varepsilon+\varepsilon'+\omega,  -\omega'_-)
\big]
\Big\}
~.
\nonumber\\
\end{eqnarray}
Note that the terms on the r.h.s. of Eq.~(\ref{eq:W_3_f_BC}) proportional to $n_{\rm B}(\omega')$ are identical in both $W^{(3,\sigma\sigma'')}_{\lambda\lambda'\mu\mu'} ({\bm k}',{\bm k},\varepsilon'_\pm-\varepsilon)$ and thus vanish when the difference is taken. We thus neglect these terms in what follows. Eq.~(\ref{eq:W_3_f_BC}) reduces to
\begin{eqnarray} \label{eq:W_3_f_final}
&& \!\!\!\!\!\!\!\!
W^{(3,\sigma\sigma'')}_{\lambda\lambda'\mu\mu'} ({\bm k}',{\bm k},\varepsilon'_\pm+\varepsilon+\omega) =
-\int \frac{d\omega'}{2\pi i} \Big\{
n_{\rm F}(\omega'+\varepsilon)
\nonumber\\
&\times&
\big[w_3(\omega'_+,\omega'_--\omega,\omega'_++\varepsilon, \varepsilon'+\omega-\omega'_\mp)
\nonumber\\
&-&
w_3(\omega'_+,\omega'_--\omega,\omega'_-+\varepsilon, \varepsilon'+\omega-\omega'_\mp)
\big]
\nonumber\\
\!\!\! &+& \!\!\!
n_{\rm F}(\omega'-\varepsilon'-\omega) \big[w_3(\omega'_+,\omega'_--\omega,\omega'_\pm + \varepsilon,  \varepsilon'+\omega-\omega'_+)
\nonumber\\
&-&
w_3(\omega'_+,\omega'_--\omega,\omega'_\pm + \varepsilon,  \varepsilon'+\omega-\omega'_-)
\big]
\Big\}
~.
\nonumber\\
\end{eqnarray}
Finally, substituting Eq.~(\ref{eq:W_3}) into Eq.~(\ref{eq:W_3_f_final}) we get
\begin{eqnarray} \label{eq:W_3_A-R}
&& \!\!\!\!\!\!\!\!
W^{(3,\sigma\sigma'')}_{\lambda\lambda'\mu\mu'} (\varepsilon'_-+\varepsilon+\omega) - W^{(3,\sigma\sigma'')}_{\lambda\lambda'\mu\mu'} (\varepsilon'_++\varepsilon+\omega) =
\int \frac{d\omega'}{2\pi i} 
\nonumber\\
&\times&
\big[n_{\rm F}(\omega'+\varepsilon) - n_{\rm F}(\omega'-\varepsilon'-\omega)\big]
\nonumber\\
&\times&
\big[
w_3(\omega'_+,\omega'_--\omega,\omega'_++\varepsilon, \varepsilon'+\omega-\omega'_-)
\nonumber\\
&-&
w_3(\omega'_+,\omega'_-,\omega'_-+\varepsilon, \varepsilon'+\omega-\omega'_-)
\nonumber\\
&-&
w_3(\omega'_+,\omega'_--\omega,\omega'_++\varepsilon, \varepsilon'+\omega-\omega'_+)
\nonumber\\
&+&
w_3(\omega'_+,\omega'_--\omega,\omega'_-+\varepsilon, \varepsilon'+\omega-\omega'_+)
\big]
\nonumber\\
&=&
-4 N_{\rm v} \sum_{{\bm q}'} \sum_{\lambda'',\mu''} \int \frac{d\omega'}{2\pi i} W({\bm q}',\omega'_+) W({\bm q}',\omega'_--\omega)
\nonumber\\
&\times&
\big[n_{\rm F}(\omega'+\varepsilon) - n_{\rm F}(\omega'-\varepsilon'-\omega)\big]
\nonumber\\
&\times&
\Im m \big[ G^{({\rm R},\sigma)}_{\lambda''}({\bm k}_- +{\bm q}',\varepsilon+\omega') \big]
\nonumber\\
&\times&
\Im m \big[ G^{({\rm R},\sigma'')}_{\mu''}({\bm k}'_+ -{\bm q}',\varepsilon' +\omega- \omega') \big]
\nonumber\\
&\times&
{\cal D}_{\lambda\lambda''}({\bm k}_-,{\bm k}_-+{\bm q}')
{\cal D}_{\lambda''\lambda'}({\bm k}_-+{\bm q}',{\bm k}_+)
\nonumber\\
&\times&
{\cal D}_{\mu\mu''}({\bm k}'_-,{\bm k}'_+-{\bm q}')
{\cal D}_{\mu''\mu'}({\bm k}'_+-{\bm q}',{\bm k}'_+)
~.
\end{eqnarray}
Taking the limit $v_{\rm F} q \ll \omega\ll \varepsilon_{\rm F}$, Eq.~(\ref{eq:W_3_A-R}) becomes
\begin{eqnarray}\label{eq:W_3_A-R_2}
&&
W^{(3,\sigma\sigma'')}_{\lambda\lambda'\mu\mu'} (\varepsilon'_-+\varepsilon) - W^{(3,\sigma\sigma'')}_{\lambda\lambda'\mu\mu'} (\varepsilon'_++\varepsilon) =
-4 N_{\rm v} \sum_{{\bm q}',\sigma''} 
\nonumber\\
&& \times
\sum_{\lambda'',\mu''} \int \frac{d\omega'}{2\pi i} \big| W({\bm q}',\omega')\big|^2
\big[n_{\rm F}(\omega'+\varepsilon) - n_{\rm F}(\omega'-\varepsilon')\big]
\nonumber\\
&& \times
\Im m \big[ G^{({\rm R},\sigma)}_{\lambda''} ({\bm k} +{\bm q}',\varepsilon+\omega') \big]
\nonumber\\
&& \times
\Im m \big[ G^{({\rm R},\sigma'')}_{\mu''} ({\bm k}' -{\bm q}',\varepsilon' - \omega') \big]
\nonumber\\
&& \times
{\cal D}_{\lambda\lambda''}({\bm k},{\bm k}+{\bm q}')
{\cal D}_{\lambda''\lambda'}({\bm k}+{\bm q}',{\bm k})
\nonumber\\
&& \times
{\cal D}_{\mu\mu''}({\bm k}',{\bm k}'-{\bm q}')
{\cal D}_{\mu''\mu'}({\bm k}'-{\bm q}',{\bm k}')
~.
\nonumber\\
\end{eqnarray}

\section{The derivation of the Bethe-Salpeter equation in the charge and spin channel} \label{app:BS_spin}
Using Eq.~(\ref{eq:S_R_GR_GA_product}) in Eqs.~(\ref{eq:R_Lambda_1_together})-(\ref{eq:R_Lambda_3_together}) we get
\begin{eqnarray} \label{eq:S_R_Lambda_1_together}
&& \!\!\!\!\!\!\!\!
\Lambda^{(1,\sigma\sigma')}_{\lambda\lambda,\beta} ({\bm k}, \varepsilon_++\omega,{\bm k},\varepsilon_-) =
-\frac{8 i N_{\rm v}}{\omega + i/\tau_{\rm ee}}
\sum_{{\bm k}',{\bm q}',\sigma''} \sum_{\mu, \mu'',\lambda''}
\nonumber\\
&\times&
\int \frac{d\varepsilon'}{2\pi i} \int \frac{d\omega'}{2\pi i}
\big[n_{\rm F}(\varepsilon') + n_{\rm B}(\varepsilon'-\varepsilon)\big]
\nonumber\\
&\times&
\big[n_{\rm F} (\omega'+\varepsilon') - n_{\rm F} (\omega'+\varepsilon)\big] 
|W({\bm k}-{\bm k}',\varepsilon'-\varepsilon)|^2
\nonumber\\
&\times&
\Im m \big[G^{({\rm R},\sigma'')}_{\lambda''}({\bm q}'-{\bm k},\omega'+\varepsilon) \big]
\Im m\big[ G^{({\rm R},\sigma)}_{\mu}({\bm k}', \varepsilon') \big]
\nonumber\\
&\times&
\Im m\big[G^{({\rm R},\sigma'')}_{\mu''}({\bm q'}-{\bm k}', \omega'+\varepsilon')\big]
{\cal D}_{\lambda\mu}({\bm k},{\bm k}') 
\nonumber\\
&\times&
{\cal D}_{\mu\lambda}({\bm k}',{\bm k})
{\cal D}_{\lambda''\mu''}({\bm q}'-{\bm k},{\bm q}'-{\bm k}')
\nonumber\\
&\times&
{\cal D}_{\mu''\lambda''}({\bm q}'-{\bm k}',{\bm q}'-{\bm k}) 
\Lambda^{(\sigma\sigma')}_{\mu\mu,\beta} ({\bm k}', \varepsilon'_++\omega, {\bm k}',\varepsilon'_-)
~,
\nonumber\\
\end{eqnarray}
and
\begin{eqnarray} \label{eq:S_R_Lambda_2_together}
&& \!\!\!\!\!\!\!\!
\Lambda^{(2,\sigma\sigma')}_{\lambda\lambda,\beta} ({\bm k}, \varepsilon_++\omega,{\bm k},\varepsilon_-) =
-\frac{8 i N_{\rm v}}{\omega + i/\tau_{\rm ee}} \sum_{{\bm k}',{\bm q}',\sigma''} \sum_{\mu,\lambda'',\mu''}
\nonumber\\
&\times&
\int \frac{d\varepsilon'}{2\pi i} \int \frac{d\omega'}{2\pi i} 
\big[n_{\rm F}(\varepsilon') + n_{\rm B}(\varepsilon'-\varepsilon)\big]
\nonumber\\
&\times&
\big[n_{\rm F}(\omega'-\varepsilon) - n_{\rm F}(\omega'-\varepsilon')\big]
|W({\bm q}',\omega')|^2 
\nonumber\\
&\times&
\Im m\big[ G^{({\rm R},\sigma'')}_{\mu''}({\bm k}'-{\bm q}',\varepsilon' -\omega') \big]
\Im m\big[ G^{({\rm R},\sigma'')}_{\mu}({\bm k}', \varepsilon') \big]
\nonumber\\
&\times&
\Im m\big[ G^{({\rm R},\sigma)}_{\lambda''}({\bm k}-{\bm q}',\varepsilon-\omega') \big]
{\cal D}_{\lambda\lambda''}({\bm k},{\bm k}-{\bm q}')
\nonumber\\
&\times&
{\cal D}_{\lambda''\lambda}({\bm k}-{\bm q}',{\bm k})
{\cal D}_{\mu\mu''}({\bm k}',{\bm k}'-{\bm q}')
{\cal D}_{\mu''\mu}({\bm k}'-{\bm q}',{\bm k}')
\nonumber\\
&\times&
\Lambda^{(\sigma''\sigma')}_{\mu\mu,\beta} ({\bm k}', \varepsilon'_++\omega, {\bm k}',\varepsilon'_-)
~,
\end{eqnarray}
and finally
\begin{eqnarray} \label{eq:S_R_Lambda_3_together}
&& \!\!\!\!\!\!\!\!
\Lambda^{(3,\sigma\sigma')}_{\lambda\lambda,\beta} ({\bm k}, \varepsilon_++\omega,{\bm k},\varepsilon_-) =
\frac{8 i N_{\rm v}}{\omega + i/\tau_{\rm ee}} \sum_{{\bm k}',{\bm q}',\sigma''} \sum_{\mu,\lambda'',\mu''}
\nonumber\\
&\times&
\int \frac{d\varepsilon'}{2\pi i} \int \frac{d\omega'}{2\pi i} 
\big[n_{\rm F}(\varepsilon') + n_{\rm B}(\varepsilon'+\varepsilon)\big]
\nonumber\\
&\times&
\big[n_{\rm F}(\omega'+\varepsilon) - n_{\rm F}(\omega'-\varepsilon')\big]
|W({\bm q}',\omega')|^2
\nonumber\\
&\times&
\Im m \big[ G^{({\rm R},\sigma)}_{\lambda''}({\bm k} +{\bm q}',\omega'+\varepsilon) \big]
\Im m \big[ G^{({\rm R},\sigma'')}_{\mu}({\bm k}',\varepsilon')]
\nonumber\\
&\times&
\Im m \big[ G^{({\rm R},\sigma'')}_{\mu''}({\bm k}' -{\bm q}',\varepsilon'- \omega') \big]
{\cal D}_{\lambda\lambda''}({\bm k},{\bm k}+{\bm q}')
\nonumber\\
&\times&
{\cal D}_{\lambda''\lambda}({\bm k}+{\bm q}',{\bm k})
{\cal D}_{\mu\mu''}({\bm k}',{\bm k}'-{\bm q}')
{\cal D}_{\mu''\mu}({\bm k}'-{\bm q}',{\bm k}')
\nonumber\\
&\times&
\Lambda^{(\sigma''\sigma')}_{\mu\mu,\beta} ({\bm k}', \varepsilon'_++\omega,{\bm k}',\varepsilon'_-)
~.
\end{eqnarray}
In these equation the limit $\omega\to 0$ is understood.
Eqs.~(\ref{eq:S_R_Lambda_1_together})-(\ref{eq:S_R_Lambda_3_together}) should be plugged into Eq.~(\ref{eq:R_Lambda_def}) and then back into Eq.~(\ref{eq:chi_jj_final_omega_finite}) taken in the limit $\omega \to 0$. The latter is given in Eq.~(\ref{eq:S_R_chi_jj_final_omega0}), and can be further approximated as
\begin{eqnarray} \label{eq:chi_jj_final_omega0}
&& \!\!\!\!\!\!\!\!
\chi_{j^{(\sigma)}_{\alpha} j^{(\sigma')}_{\beta}} ({\bm q}={\bm 0}, \omega) \to - \omega \frac{2 i N_{\rm v}}{\omega + i\tau_{\rm ee}} \sum_{{\bm k}} \int \frac{d\varepsilon}{2\pi i} \frac{\partial n_{\rm F} (\varepsilon)}{\partial \varepsilon}
\nonumber\\
&\times&
\Im m \big[G^{({\rm R},\sigma)}_{+}({\bm k}, 0) \big]
\Lambda^{(0,\sigma)}_{++,\alpha}({\bm k},{\bm k})
\Lambda^{(\sigma\sigma')}_{++,\beta} ({\bm k}, \varepsilon_++\omega,{\bm k},\varepsilon_-)
~.
\nonumber\\
\end{eqnarray}
To obtain this expression we used Eq.~(\ref{eq:S_R_GR_GA_product}), together with the fact that the derivative of the Fermi function is peaked in $\varepsilon=0$, and that $\Im m \big[G^{({\rm R},\sigma)}_{\lambda}({\bm k}, 0)\big]$ constrains $|{\bm k}| = k_{\rm F}$ and $\lambda=+$. We now approximate Eqs.~(\ref{eq:S_R_Lambda_1_together})-(\ref{eq:S_R_Lambda_3_together}) by noting that the combination of Fermi and Bose functions constrains $\varepsilon \sim \varepsilon' \sim \omega' \sim 0$. We thus get
\begin{eqnarray} \label{eq:S_R_Lambda_1_together_2}
&& \!\!\!\!\!\!\!\!
\Lambda^{(1,\sigma\sigma')}_{++,\beta} ({\bm k}, \varepsilon_++\omega,{\bm k},\varepsilon_-) =
-\frac{8 i N_{\rm v}}{\omega + i/\tau_{\rm ee}}
\sum_{{\bm k}',{\bm q}',\sigma''}
\nonumber\\
&\times&
\int \frac{d\varepsilon'}{2\pi i} \int \frac{d\omega'}{2\pi i}
\big[n_{\rm F}(\varepsilon') + n_{\rm B}(\varepsilon'-\varepsilon)\big]
\nonumber\\
&\times&
\big[n_{\rm F} (\omega'+\varepsilon') - n_{\rm F} (\omega'+\varepsilon)\big] 
|W({\bm q}',0)|^2
\nonumber\\
&\times&
\Im m \big[ G^{({\rm R},\sigma)}_{+}({\bm k}-{\bm q}',0) \big]
\Im m \big[ G^{({\rm R},\sigma'')}_{+}({\bm k}',0) \big]
\nonumber\\
&\times&
\Im m\big[G^{({\rm R},\sigma'')}_{+}({\bm q}'-{\bm k}', 0) \big]
{\cal D}_{++}({\bm k},{\bm k}-{\bm q}')
{\cal D}_{++}({\bm k}-{\bm q}',{\bm k})
\nonumber\\
&\times&
{\cal D}_{++}({\bm k}',{\bm k}'-{\bm q}')
{\cal D}_{++}({\bm k}'-{\bm q}',{\bm k}') 
\nonumber\\
&\times&
\Lambda^{(\sigma\sigma')}_{++,\beta} ({\bm k}-{\bm q}', \omega^+, {\bm k}-{\bm q}',0^-)
~,
\nonumber\\
\end{eqnarray}
and
\begin{eqnarray} \label{eq:S_R_Lambda_2_together_2}
&& \!\!\!\!\!\!\!\!
\Lambda^{(2,\sigma\sigma')}_{++,\beta} ({\bm k}, \varepsilon_++\omega,{\bm k},\varepsilon_-) =
-\frac{8 i N_{\rm v}}{\omega + i/\tau_{\rm ee}} \sum_{{\bm k}',{\bm q}',\sigma''}
\nonumber\\
&\times&
\int \frac{d\varepsilon'}{2\pi i} \int \frac{d\omega'}{2\pi i} 
\big[n_{\rm F}(\varepsilon') + n_{\rm B}(\varepsilon'-\varepsilon)\big]
\nonumber\\
&\times&
\big[n_{\rm F}(\omega'+\varepsilon') - n_{\rm F}(\omega'+\varepsilon)\big]
|W({\bm q}',0)|^2 
\nonumber\\
&\times&
\Im m\big[ G^{({\rm R},\sigma'')}_{+}({\bm k}'-{\bm q}',0) \big]
\Im m\big[ G^{({\rm R},\sigma'')}_{+}({\bm k}',0) \big]
\nonumber\\
&\times&
\Im m\big[ G^{({\rm R},\sigma)}_{+}({\bm k}-{\bm q}',0) \big]
{\cal D}_{++}({\bm k},{\bm k}-{\bm q}')
\nonumber\\
&\times&
{\cal D}_{++}({\bm k}-{\bm q}',{\bm k})
{\cal D}_{++}({\bm k}',{\bm k}'-{\bm q}')
{\cal D}_{++}({\bm k}'-{\bm q}',{\bm k}')
\nonumber\\
&\times&
\Lambda^{(\sigma''\sigma')}_{++,\beta} ({\bm k}', \omega^+, {\bm k}',0^-)
~,
\end{eqnarray}
and finally
\begin{eqnarray} \label{eq:S_R_Lambda_3_together_2}
&& \!\!\!\!\!\!\!\!
\Lambda^{(3,\sigma\sigma')}_{++,\beta} ({\bm k}, \varepsilon_++\omega,{\bm k},\varepsilon_-) =
\frac{8 i N_{\rm v}}{\omega + i/\tau_{\rm ee}} \sum_{{\bm k}',{\bm q}',\sigma''} 
\nonumber\\
&\times&
\int \frac{d\varepsilon'}{2\pi i} \int \frac{d\omega'}{2\pi i} 
\big[n_{\rm F}(\varepsilon') + n_{\rm B}(\varepsilon'-\varepsilon)\big]
\nonumber\\
&\times&
\big[n_{\rm F}(\omega'+\varepsilon') - n_{\rm F}(\omega'+\varepsilon)\big]
|W({\bm q}',0)|^2
\nonumber\\
&\times&
\Im m \big[ G^{({\rm R},\sigma)}_{+}({\bm k} -{\bm q}',0) \big]
\Im m \big[ G^{({\rm R},\sigma'')}_{+}({\bm k}'-{\bm q}',0)]
\nonumber\\
&\times&
\Im m \big[ G^{({\rm R},\sigma'')}_{+}({\bm k}',0) \big]
{\cal D}_{++}({\bm k},{\bm k}-{\bm q}')
\nonumber\\
&\times&
{\cal D}_{++}({\bm k}-{\bm q}',{\bm k})
{\cal D}_{++}({\bm k}'-{\bm q}',{\bm k}')
{\cal D}_{++}({\bm k}',{\bm k}'-{\bm q}')
\nonumber\\
&\times&
\Lambda^{(\sigma''\sigma')}_{++,\beta} ({\bm k}'-{\bm q}', \omega^+,{\bm k}'-{\bm q}',0^-)
~.
\end{eqnarray}
We shifted ${\bm k}'\to {\bm k}-{\bm q}'$ and ${\bm q}'\to {\bm k}-{\bm k}'$ in Eq.~(\ref{eq:S_R_Lambda_1_together_2}), $\omega' \to -\omega'$ in Eq.~(\ref{eq:S_R_Lambda_2_together_2}), and $\varepsilon' \to -\varepsilon'$, ${\bm k}'\to {\bm k}'-{\bm q}'$ and ${\bm q}' \to -{\bm q}'$ in Eq.~(\ref{eq:S_R_Lambda_3_together_2}). Moreover, we used that the imaginary parts of the Green's functions constrain their momentum argument to the Fermi surface and the band indices to be all equal to ``$+$''. 
Putting everything together
\begin{eqnarray} \label{eq:S_R_Lambda_all_together}
&& \!\!\!\!\!\!\!\!
\sum_{i=1}^3 \Lambda^{(i,\sigma\sigma')}_{++,\beta} ({\bm k}, \omega^+,{\bm k},0^-) =
-\frac{8 i N_{\rm v}}{\omega + i/\tau_{\rm ee}}
\sum_{{\bm k}',{\bm q}',\sigma''}|W({\bm q}',0)|^2
\nonumber\\
&\times&
\int \frac{d\varepsilon'}{2\pi i} \int \frac{d\omega'}{2\pi i}
\big[n_{\rm F}(\varepsilon') + n_{\rm B}(\varepsilon'-\varepsilon)\big]
\nonumber\\
&\times&
\big[n_{\rm F} (\omega'+\varepsilon') - n_{\rm F} (\omega'+\varepsilon)\big] 
\Im m \big[ G^{({\rm R},\sigma'')}_{+}({\bm k}',0) \big]
\nonumber\\
&\times&
\Im m \big[ G^{({\rm R},\sigma)}_{+}({\bm k}-{\bm q}',0) \big]
\Im m\big[G^{({\rm R},\sigma'')}_{+}({\bm q}'-{\bm k}', 0) \big]
\nonumber\\
&\times&
{\cal D}_{++}({\bm k},{\bm k}-{\bm q}')
{\cal D}_{++}({\bm k}-{\bm q}',{\bm k})
\nonumber\\
&\times&
{\cal D}_{++}({\bm k}',{\bm k}'-{\bm q}')
{\cal D}_{++}({\bm k}'-{\bm q}',{\bm k}') 
\nonumber\\
&\times&
\Big[
\Lambda^{(\sigma\sigma')}_{++,\beta} ({\bm k}-{\bm q}', \omega^+, {\bm k}-{\bm q}',0^-)
+ \Lambda^{(\sigma''\sigma')}_{++,\beta} ({\bm k}', \omega^+, {\bm k}',0^-)
\nonumber\\
&-&
\Lambda^{(\sigma''\sigma')}_{++,\beta} ({\bm k}'-{\bm q}', \omega^+,{\bm k}'-{\bm q}',0^-)
\Big]
~.
\end{eqnarray}
Plugging Eq.~(\ref{eq:S_R_Lambda_all_together}) back into Eq.~(\ref{eq:R_Lambda_def}) we immediately get the self-consistent Bethe-Salpeter Eq.~(\ref{eq:S_Bethe_Salpeter_final}).

\section{The spin transport time}
\label{app:spin_transport_time}
The spin transport time defined in Eq.~(\ref{eq:S_tau_tr}), which we recall here for completeness, reads
\begin{eqnarray} \label{eq:S_tau_tr_R}
\frac{1}{\tau^{({\rm s})}_{\rm tr}}  &=& 
- \frac{8}{3}  N_{\rm v} (k_{\rm B} T)^2
\sum_{{\bm k}',{\bm q}'}|W({\bm q}',0)|^2
\Im m \big[ G^{({\rm R})}_{+}({\bm k}-{\bm q}',0) \big]
\nonumber\\
&\times&
\Im m \big[ G^{({\rm R})}_{+}({\bm k}',0) \big]
\Im m\big[G^{({\rm R})}_{+}({\bm q}'-{\bm k}', 0) \big]
\nonumber\\
&\times&
{\cal D}_{++}({\bm k},{\bm k}-{\bm q}')
{\cal D}_{++}({\bm k}-{\bm q}',{\bm k})
\nonumber\\
&\times&
{\cal D}_{++}({\bm k}',{\bm k}'-{\bm q}')
{\cal D}_{++}({\bm k}'-{\bm q}',{\bm k}') 
\nonumber\\
&\times&
\big[1 - \cos(\varphi_{\bm k} - \varphi_{{\bm k}-{\bm q}'})\big]
~.
\end{eqnarray}
In the low-temperature limit the three Green's functions on the right-hand side of Eq.~(\ref{eq:S_tau_tr_R}) can be approximated by $\delta$-functions, which constrain
\begin{equation} \label{eq:app_deltas_solutions}
\left\{
\begin{array}{l}
{\displaystyle \cos(\varphi_{{\bm k}}-\varphi_{{\bm q}'}) = \frac{q'}{2 k_{\rm F}} }
\vspace{0.3cm}\\
{\displaystyle \cos(\varphi_{{\bm k}'}-\varphi_{{\bm q}'}) = \frac{q'}{2 k_{\rm F}} }
\end{array}
\right.
~.
\end{equation}
and thus $1 - \cos(\varphi_{\bm k} - \varphi_{{\bm k}-{\bm q}'}) = q'^2/(2k_{\rm F}^2)$. The Coulomb interaction forces us to disregard the solution of the three $\delta$-functions with $q'=0$. 

Note that Eq.~(\ref{eq:S_tau_tr_R}) describes the simultaneous scattering of a particle from the state ${\bm k}$ to the state ${\bm k}-{\bm q}'$, both at the Fermi surface, and the creation of a particle-hole pair with total momentum ${\bm q}'$. According to Fig.~\ref{fig:three}, the angles between ${\bm k}$ and ${\bm k}-{\bm q}'$, and ${\bm k}'$ and ${\bm k}'-{\bm q}'$, are identical. This implies that ${\cal D}_{++}({\bm k},{\bm k}-{\bm q}') = {\cal D}_{++}({\bm k}',{\bm k}'-{\bm q}')$. Moreover,
\begin{eqnarray}
{\cal D}_{++}({\bm k},{\bm k}-{\bm q}') {\cal D}_{++}({\bm k}-{\bm q}',{\bm k}) = 1 - \left(\frac{q'}{2k_{\rm F}}\right)^2
~.
\end{eqnarray}
Putting everything together we get
\begin{eqnarray} \label{eq:app_S_tau_tr_1}
\frac{1}{\tau^{({\rm s})}_{\rm tr}}  &=& 
- \frac{16}{3}  N_{\rm v} (k_{\rm B} T)^2
\sum_{{\bm k}',{\bm q}'}|W({\bm q}',0)|^2
\Im m \big[ G^{({\rm R})}_{+}({\bm k}-{\bm q}',0) \big]
\nonumber\\
&\times&
\Im m \big[ G^{({\rm R})}_{+}({\bm k}',0) \big]
\Im m\big[G^{({\rm R})}_{+}({\bm q}'-{\bm k}', 0) \big]
\nonumber\\
&\times&
\left[1 - \left(\frac{q'}{2k_{\rm F}}\right)^2 \right]^2
\left(\frac{q'}{2k_{\rm F}}\right)^2
~.
\end{eqnarray}
Shifting $\varphi_{{\bm k}'} \to \varphi_{{\bm k}'} + \varphi_{{\bm k}}$ and $\varphi_{{\bm q}'} \to \varphi_{{\bm q}'} + \varphi_{{\bm k}}$, we immediately see that the three $\delta$-functions are solved by $\varphi_{{\bm q}'} = \pm \arccos\big[q'/(2 k_{\rm F})\big]$ and $\varphi_{{\bm k}'} = 0, 2\varphi_{{\bm q}'}$. All these solutions give identical contributions. Summing all of them we finally get
\begin{eqnarray} \label{eq:app_S_tau_tr_final}
\frac{1}{\tau^{({\rm s})}_{\rm tr}}  &=& 
\frac{N_{\rm v} (k_{\rm B} T)^2}{3\pi v_{\rm F}^2 \varepsilon_{\rm F}}  
\int_0^{2 k_{\rm F}} dq' ~q' |W({\bm q}',0)|^2
\nonumber\\
&\times&
\left[1 - \left(\frac{q'}{2k_{\rm F}}\right)^2 \right]
\nonumber\\
&=&
\frac{2\pi N_{\rm v}}{3} \frac{\alpha_{\rm ee}^2(k_{\rm B} T)^2}{\hbar \varepsilon_{\rm F}}
\big[ 3 (2 N_{\rm v} \alpha_{\rm ee} -1) \nonumber\\
&+&
4(1 - 3 N_{\rm v}^2 \alpha_{\rm ee}^2) {\rm arccoth}(1 + 2 N_{\rm v}\alpha_{\rm ee}) \big]
~.
\end{eqnarray}
Here we used that for $0<q'<2 k_{\rm F}$ $W({\bm q}',0)$ reduces to the Thomas-Fermi interaction, with $q_{\rm TF} = 2 N_{\rm v} \alpha_{\rm ee} k_{\rm F}$ as the screening wavevector. In the limit $\alpha_{\rm ee}\to 0$ we get
\begin{eqnarray} \label{eq:app_S_tau_tr_final_limit}
\frac{1}{\tau^{({\rm s})}_{\rm tr}}  &\to& 
- \frac{4\pi N_{\rm v}}{3} \frac{(k_{\rm B} T)^2}{\hbar \varepsilon_{\rm F}} \alpha_{\rm ee}^2 \ln(N_{\rm v} \alpha_{\rm ee})
~.
\end{eqnarray}
The logarithmic dependence on the coupling constant $\alpha_{\rm ee}$ is due to the Thomas-Fermi screening.

\section{The spin velocity}
\label{eq:spin_velocity}
We now show a brief derivation of the renormalization of the spin Drude weight of Eq.~(\ref{eq:spin_drude_weight}), based on Landau theory of normal Fermi liquids. The derivation closely follows that of Ref.~\onlinecite{Qian_prl_2004}. The question we answer in this appendix is: what is the spin current carried by a quasiparticle? We thus consider a state in which a quasiparticle with momentum ${\bm p}$ and spin up is added to the system. The adiabatically turned-on electron-electron interactions dress the quasiparticle and renormalize the spin current it carries. We consider the expectation value of the spin current operator ${\hat {\bm j}}_{\rm s} = {\hat {\bm j}}_\uparrow - {\hat {\bm j}}_\downarrow$ on this state. In analogy with the charge current, we know that the spin current is proportional to the unit vector ${\hat {\bm p}}$. However, the constant of proportionality, namely the spin velocity $v_{\rm s}$, is to be determined. We define $v_{\rm s}$ as
\begin{equation}
\langle {\bm p}, \uparrow | {\hat {\bm j}}_{\rm s} | {\bm p}, \uparrow \rangle = v_{\rm s} {\hat {\bm p}}
~.
\end{equation}
where $| {\bm p}, \uparrow \rangle$ denotes the full many-body state with the extra quasiparticle with momentum ${\bm p}$ and spin up.

We now connect the phenomenological theory of Landau to the microscopic model. Since we are interested in properties at the Fermi surface, we will consider in what follows the following one band model of graphene
\begin{equation} \label{eq:Hamiltonian_one_band}
{\hat {\cal H}} = \sum_{i} v_{\rm F} p_i + \frac{1}{2} \sum_{j\neq i} V({\bm r}_i - {\bm r}_j)
~,
\end{equation}
where $i$ and $j$ label particles in conduction band. The interaction with states in the valence band provides the well-known logarithmic divergence~\cite{Gonzalez_nuclphys_1994} of the renormalized Fermi velocity $v_{\rm F}^\star$. We now imagine to perform the following unitary transformation
\begin{equation} \label{eq:app_unitary_transform}
{\hat U} = \exp\Bigg[ i \sum_i {\hat \tau}_z {\bm q} \cdot {\hat {\bm r}}_i \Bigg]
~,
\end{equation}
on the Hamiltonian of Eq.~(\ref{eq:Hamiltonian_one_band}). In Eq.~(\ref{eq:app_unitary_transform}) we introduced the Pauli matrix ${\hat \tau}_z$ which acts on the spin degree of freedom. Clearly the transformation ${\hat U}$ commutes with the interaction Hamiltonian, since it contains only the position operator ${\hat {\bm r}}_i$. To first order in ${\bm q}$ it induces the change in energy $\Delta E = {\hat {\bm j}}_{\rm s}\cdot{\bm q}$. For the state $| {\bm p}, \uparrow \rangle$, with the extra quasiparticle of momentum ${\bm p}$ and spin up we thus get
\begin{equation} \label{eq:app_Delta_E_1}
\Delta E = v_{\rm s} {\hat {\bm p}}\cdot {\bm q}
~.
\end{equation}

On the other hand, we may consider the variation of energy due to the shift generated by the unitary transformation of Eq.~(\ref{eq:app_unitary_transform}) on the phenomenological energy functional of Landau's theory of normal Fermi liquids~\cite{Giuliani_and_Vignale}. Shifting the momentum of spin up (down) particles by ${\bm q}$ ($-{\bm q}$) we get
\begin{eqnarray} \label{eq:app_Delta_E_2}
\Delta E &=& v_{\rm F}^\star {\hat {\bm p}}\cdot{\bm q} - \sum_{{\bm p}',\tau} f_{{\bm p}\uparrow,{\bm p}'\tau} \tau {\bm q}\cdot{\bm \nabla}_{{\bm p}'} n_{0,\tau}({\bm p}')
\nonumber\\
&=&
v_{\rm F}^\star (1 + F_1^{\rm a}) {\hat {\bm p}}\cdot{\bm q}
~,
\end{eqnarray}
where $f_{{\bm p}\sigma,{\bm p}'\tau}$ is the Landau interaction function~\cite{Giuliani_and_Vignale} and $n_{0,\tau}({\bm p}')$ is the equilibrium distribution function of quasiparticles. Note that at, if ${\bm p}$ and ${\bm p}'$ are both at the Fermi surface $f_{{\bm p}\sigma,{\bm p}'\tau}$ is a function only of the difference between the angles of ${\bm p}$ and ${\bm p}'$, {\it i.e.} $f_{{\bm p}\sigma,{\bm p}'\tau} = f_{\sigma\tau}(\varphi_{{\bm p}}-\varphi_{{\bm p}'})$. The Landau parameters $F_n^{\ell}$, with $\ell={\rm a}, {\rm s}$, are defined as~\cite{Giuliani_and_Vignale}
\begin{equation} \label{eq:app_Landau_parameters}
F_n^{\ell} = \frac{k_{\rm F}}{2\pi v_{\rm F}^\star} \int \frac{d\theta}{2\pi} \big[ f_{\uparrow \uparrow}(\theta) \pm f_{\uparrow \downarrow}(\theta) \big] \cos^2 (\theta)
~,
\end{equation}
where the plus (minus) sign in square brackets on the right-hand side of Eq.~(\ref{eq:app_Landau_parameters}) holds for $\ell = {\rm s}$ ($\ell ={\rm a}$). 

A direct comparison of Eq.~(\ref{eq:app_Delta_E_1}) with Eq.~(\ref{eq:app_Delta_E_2}) immediately gives
\begin{equation}
v_{\rm s} = v_{\rm F}^\star (1 + F_1^{\rm a})
~.
\end{equation}
We obtain the renormalization of the spin Drude weight by replacing $v_{\rm s}$ in lieu of the bare Fermi velocity $v_{\rm F}$ in its non-interacting expression. 

\end{document}